\documentclass[10.5pt,dvips]{article}
\usepackage{float}
\usepackage{graphicx}
\usepackage{amsmath}
\usepackage{amssymb}
\usepackage{latexsym}
\usepackage{color}

\begin{document}
\thispagestyle{empty}
\begin{center}

\vspace{1.8cm}

 {\bf The dynamics of  local quantum uncertainty and trace distance discord for two-qubit $X$ states under decoherence: A comparative study}\\

\vspace{1.5cm}

{\bf A. Slaoui}$^{a}${\footnote { email: {\sf
abdallahsalaoui1992@gmail.com}}}, {\bf M. Daoud}$^{b}${\footnote {
email: {\sf m$_{-}$daoud@hotmail.com}}} and {\bf R. Ahl
Laamara}$^{a,c}$ {\footnote { email: {\sf ahllaamara@gmail.com}}}

\vspace{0.5cm}

$^{a}${\it LPHE-Modeling and Simulation, Faculty  of Sciences,
University
Mohammed V,\\ Rabat, Morocco}\\[1em]

$^{b}${\it Department of Physics , Faculty of Sciences-Ain Chock, University Hassan II,\\
 Casablanca,
Morocco}\\[1em]

$^{c}${\it Centre of Physics and Mathematics,
CPM, CNESTEN,\\ Rabat, Morocco}\\[1em]

\vspace{3cm} {\bf Abstract}
\end{center}
\baselineskip=18pt
\medskip

We employ the concepts of local quantum uncertainty and geometric  quantum discord based on the trace norm to investigate the environmental effects on  quantum correlations of
two bipartite quantum systems. The first one concerns a two-qubit system coupled with two independent bosonic reservoirs. We show that the trace discord exhibits frozen phenomenon contrarily to local quantum uncertainty.  The second
scenario deals with a two level system, initially prepared in a separable state, interacting with a quantized electromagnetic radiation.  Our results show that
there exists an exchange of quantum correlations between the two-level system and its surrounding which is responsible of the revival phenomenon
of non classical correlations. \\

Dynamics of two-qubit system. Frozen and revival quantum correlations.

\newpage

\section{Introduction}

Quantum information protocols exploit the quantum features of superposition and entanglement to achieve quantum tasks that
are not possible using the classical laws of physics and  quantum information processors offer significant advantages in the
communication and  processing of information \cite{Nielsen}.  Quantum teleportation \cite{Braunstein,Bouwmeester},
 and quantum cryptography or quantum key  distribution \cite{BB84} constitute the most promising  applications of quantum information science. Characterizing quantum correlations in quantum systems is a challenging issue in this field of research. Several measures of quantum correlations in bipartite and multipartite quantum systems
have been introduced in the literature \cite{Vedral-RMP-2002, Horodecki-RMP-2009,Guhne, vedral} . The concurrence and entanglement of formation are two examples of quantitative measures of entanglement \cite{Wootters}.  Another indicator
of quantum correlations in bipartite quantum systems is the quantum discord based on the von Neumann entropy. This measure was introduced Ollivier and Zurek \cite{Ollivier2001} and by Henderson and Vedral \cite{Henderson2001}
and goes beyond entanglement and provides the proper tool to investigate the quantum correlations in an arbitrary bipartite state even those which are separable. The computability of quantum discord based on von Neumann entropy is in general a very complex task. Indeed, it has been proven that computing entropic quantum discord is NP-complete \cite{Huang2014} and only partial results were obtained for some special two-qubit states. To overcome these difficulties, a geometric variant of quantum discord has been proposed in  several works by employing Schatten $p$-norms. The first geometric measure
 of quantum discord was formulated in  \cite{Dakic2010} by using the Hilbert-Schmidt norm $(p = 2)$. Despite its ease of computability \cite{Bellomo1,Bellomo2,DaoudPLA,DaoudIJQI}, the geometric discord
 is not a good quantum correlations indicator. In fact, the geometric discord based on Hilbert-Schmidt distance can increase under local quantum operations on the unmeasured qubit. This drawback is due to the lack of the contractivity property that any quantum correlations quantifier should satisfy (see reference \cite{piani}). Now, it is well established that  the trace norm ( Bures norm with $p=1$) is the
  only Schatten $p$-norm  that one can use to deal with the geometric measure of quantum correlations \cite{paula} (see also \cite{Bromley,Ciccarello}).

Recently, a new measure called local quantum
uncertainty has been introduced to investigate the pairwise quantum
correlations of the discord type in multipartite systems. It has all
the desirable properties that every good  quantum correlations
quantifier should satisfy \cite{girolami}. It is based on the
notion of Skew information introduced by Wigner to determine the
uncertainty in the measurement of an observable \cite{wigner}. The
local quantum uncertainty presents the advantage of being
analytically computable for any qubit-qudit system \cite{girolami}.
It is interesting to stress that this new quantum
correlation quantifier is deeply related to quantum Fisher
information \cite{luo,petz,luo2} which is usually used in the
context of quantum metrology \cite{Girolami:13}.

In this paper, to quantify the degree of quantum correlations, we will use the local quantum uncertainty and the trace quantum discord. In Section 2, we give the explicit expression of local quantum uncertainty
in a two-qubit $X$ state. We give also the expression of the geometric discord based on the trace norm.   In  Section 3, we will investigate the dynamics of quantum correlations in a two-qubit $X$ state coupled to two independent reservoirs. Section 4 is devoted to the analysis of the creation of quantum correlations in a system of two 2-level atoms, interacting with a quantized radiation field, initially
prepared in a separable state.  Concluding remarks close this paper.

\section{Quantum correlations quantifiers: local quantum uncertainty and trace distance}

\subsection{Local quantum uncertainty}
The local quantum uncertainty is a reliable  quantifier
 of quantum correlation in bipartite quantum   systems. This is essentially due to
its easiness of computability and the fact that it enjoys all necessary properties of being a
good quantumness measure. It is zero for classically correlated states and invariant  under local unitary operations. We notice that for a two-qubit pure state, the local quantum uncertainty reduces to the linear entropy of entanglement \cite{girolami}. The local quantum uncertainty quantifies the minimal quantum uncertainty in a quantum state due to a measurement of a local
observable \cite{girolami}. For a bipartite quantum state $\rho$, the
local quantum uncertainty is defined as
\begin{equation}
\mathcal{U}(\rho) \equiv \min_{K_1} \mathcal{I}(\rho, K_1 \otimes
\mathbb{I}_2), \label{LQU}
\end{equation}
where $K_1$ is some local observable on the qubit $1$, $\mathbb{I}_2$ is the identity operator acting on the qubit $2$ and
\begin{equation}
\mathcal{I}(\rho,   K_1 \otimes
\mathbb{I}_2)=-\frac{1}{2}{\rm
Tr}([\sqrt{\rho}, K_1 \otimes
\mathbb{I}_2]^{2})
\end{equation}
 is the skew information which
provides an analytical tool to quantify the information content in the state  $\rho$ with respect to the
observable $K_1$. This quantity was introduced by Wigner and Yanase to quantify the uncertainty in
 mixed states \cite{wigner}. The statistical idea underlying skew information is the Fisher information which plays a central role in the
theory of statistical estimation and quantum metrology  \cite{luo}. The local quantum uncertainty is defined through a minimization procedure over
the ensemble of all Hermitian operators acting on the qubit 1 \cite{girolami}.  For  two qubit systems
($\frac{1}{2}$-spin particles), the expression of the local quantum uncertainty is given by
\cite{girolami}
\begin{equation}
 \mathcal{U}(\rho) = 1 - {\rm max}\{ \lambda_1, \lambda_2, \lambda_3\},
\end{equation}
where $\lambda_{1}, \lambda_2$ and $\lambda_3$ are the eigenvalues of the $3\times3$ matrix
$W$ whose matrix elements are defined by
\begin{equation}\label{w-elements}
 \omega_{ij} \equiv  {\rm
Tr}\{\sqrt{\rho}(\sigma_{i}\otimes
\mathbb{I}_2)\sqrt{\rho}(\sigma_{j}\otimes \mathbb{I}_2)\},
\end{equation}
with $i,j = 1, 2, 3$.

The $X$ states belong to an interesting family of two-qubit states that are
used in several problems of quantum information. In the computational basis $
\left\{ {\left| {00} \right\rangle ,\left| {01} \right\rangle
    ,\left| {10} \right\rangle ,\left| {11} \right\rangle } \right\}$, the $X$ states are of the form
\begin{equation}
    \rho  = \left( {\begin{array}{*{20}{c}}
            {{\rho _{11}}}&0&0&{{\rho _{14}}}\\
            0&{{\rho _{22}}}&{{\rho _{23}}}&0\\
            0&{{\rho _{32}}}&{{\rho _{33}}}&0\\
            {{\rho _{41}}}&0&0&{{\rho _{44}}}
    \end{array}} \right). \label{X}
\end{equation}
In the Fano-Bloch representation, the density matrix $\rho $ can be written as follows:
\begin{equation}
    \rho  = \frac{1}{4}\sum\limits_{\alpha ,\beta } {{T_{\alpha \beta }}} {\sigma _\alpha } \otimes {\sigma _\beta },
\end{equation}
where ${T_{\alpha \beta }} = {\rm Tr}\rho \left( {{\sigma _\alpha } \otimes {\sigma _\beta }} \right)$.  For states of $X$ type, the non vanishing matrix elements (\ref{w-elements})
are given by (see the appendix 1)
\begin{equation}\label{w11}
    {w_{11}} = \frac{1}{4}\left[ {4\left( {\sqrt {{\lambda _1}}  + \sqrt {{\lambda _4}} } \right)\left( {\sqrt {{\lambda _2}}  + \sqrt {{\lambda _3}} } \right) + \frac{{\left( {{T_{11}}^2 - {T_{22}}^2} \right) + \left( {{T_{12}}^2 - {T_{21}}^2} \right) + \left( {{T_{03}}^2 - {T_{30}}^2} \right)}}{{\left( {\sqrt {{\lambda _1}}  + \sqrt {{\lambda _4}} } \right)\left( {\sqrt {{\lambda _2}}  + \sqrt {{\lambda _3}} } \right)}}} \right],
\end{equation}
\begin{equation} \label{w22}
    {w_{22}} = \frac{1}{4}\left[ {4\left( {\sqrt {{\lambda _1}}  + \sqrt {{\lambda _4}} } \right)\left( {\sqrt {{\lambda _2}}  + \sqrt {{\lambda _3}} } \right) + \frac{{\left( {{T_{22}}^2 - {T_{11}}^2} \right) + \left( {{T_{21}}^2 - {T_{12}}^2} \right) + \left( {{T_{30}}^2 - {T_{03}}^2} \right)}}{{\left( {\sqrt {{\lambda _1}}  + \sqrt {{\lambda _4}} } \right)\left( {\sqrt {{\lambda _2}}  + \sqrt {{\lambda _3}} } \right)}}} \right],
\end{equation}
\begin{align}\label{w33}
    {w_{33}} &= \frac{1}{2}\left[ {{{\left( {\sqrt {{\lambda _1}}  + \sqrt {{\lambda _4}} } \right)}^2} + {{\left( {\sqrt {{\lambda _2}}  + \sqrt {{\lambda _3}} } \right)}^2}} \right] + \frac{1}{8}\left[ {\frac{{{{\left( {{T_{30}} + {T_{03}}} \right)}^2} - {{\left( {{T_{11}} - {T_{22}}} \right)}^2} - {{\left( {{T_{12}} + {T_{21}}} \right)}^2}}}{{{{\left( {\sqrt {{\lambda _1}}  + \sqrt {{\lambda _4}} } \right)}^2}}}} \right] \\
    & \hspace{3cm}+ \frac{1}{8}\left[ {\frac{{{{\left( {{T_{03}} - {T_{30}}} \right)}^2} - {{\left( {{T_{11}} + {T_{22}}} \right)}^2} - {{\left( {{T_{12}} - {T_{21}}} \right)}^2}}}{{{{\left( {\sqrt {{\lambda _2}}  + \sqrt {{\lambda _3}} } \right)}^2}}}} \right],
\end{align}
\begin{equation}\label{w12}
    {w_{12}} = {w_{21}}  = \frac{1}{2}\frac{{{T_{11}}{T_{21}} + {T_{22}}{T_{12}}}}{{\left( {\sqrt {{\lambda _1}}  + \sqrt {{\lambda _4}} } \right)\left( {\sqrt {{\lambda _2}}  + \sqrt {{\lambda _3}} } \right)}},
\end{equation}
\begin{equation}\label{w13}
    {w_{13}} = {w_{31}} = {w_{23}} = {w_{32}} = 0 .
\end{equation}
where $\lambda_i (i = 1, 2, 3, 4)$ are the eigenvalues of the density matrix $\rho$.
\subsection{Trace measure of geometric discord}
The trace norm (or 1-norm) was employed as a reliable geometric quantifier
 of quantum discord \cite{paula}. The expressions of trace distance quantum discord have been
analytically  derived for Bell-diagonal states
\cite{paula,Nakano}  and for an arbitrary  two-qubit $X$ state \cite{Ciccarello}. The trace distance
quantum discord for a  two-qubit state $\rho$ is defined by
\begin{eqnarray}\label{eq1}
 D_{\rm T}(\rho)= \frac{1}{2} \min_{\chi\in\Omega}||\rho-\chi||_1,
\end{eqnarray}
where the trace distance  is defined by $||\rho-\chi||_1={\rm
Tr}\sqrt{(\rho-\chi)^\dag (\rho-\chi)}$. It  measures the distance
between the state $\rho$ and  the classical-quantum  state
$\chi$ belonging to the set $\Omega$  of classical-quantum states. A generic state $\chi\in\Omega$ is of the form
$\chi = \sum_k p_k ~\Pi_{k,1}\otimes \rho_{k,2}$
where $\{p_k\}$ is a probability distribution,  $\Pi_{k,1}$ are the
orthogonal projector associated with the qubit $1$ and $\rho_{k,2}$ is
density matrix associated with  the second qubit. The phase factors $\dfrac{\rho_{14}}{|\rho_{14}|}= e^{i\theta _{14}}$ and $\dfrac{\rho_{23}}{|\rho_{23}|}= e^{i\theta _{23}}$ of the off diagonal elements can be removed using the local unitary transformations
\begin{equation*}
\left| 0 \right\rangle_1 \to \exp \left( \frac{ - i}{2}\left( \theta _{14} + \theta _{23}\right) \right)\left| 0 \right\rangle_1 \hspace{0.5 cm} \left| 0 \right\rangle_2 \to \exp \left( \frac{ - i}{2}\left( \theta _{14} - \theta _{23}\right) \right)\left| 0 \right\rangle_2 \label{transf}.
\end{equation*}
In this way, the anti-diagonal entries of the density matrix become
positive and one gets
\begin{equation*}
\rho  \to \tilde \rho  = \left( {\begin{array}{*{20}{c}}
        {{\rho _{11}}}&0&0&{\left| {{\rho _{14}}} \right|}\\
        0&{{\rho _{22}}}&{\left| {{\rho _{23}}} \right|}&0\\
        0&{\left| {{\rho _{23}}} \right|}&{{\rho _{33}}}&0\\
        {\left| {{\rho _{14}}} \right|}&0&0&{{\rho _{44}}}
        \end{array}} \right),
\end{equation*}
which rewrites in the Fano-Bloch representation as
\begin{equation*}
    \tilde \rho  = \sum\limits_{\alpha \beta } {R_{\alpha \beta }} {\sigma _\alpha } \otimes {\sigma _\beta },
\end{equation*}
where the non vanishing matrix elements $R_{\alpha \beta }$ are given by
$$R_{11}=2(|\rho_{23}|+ |\rho_{14}|) \qquad R_{22}=2(|\rho_{23}|- |\rho_{14}|) \qquad R_{33}=1-2(\rho_{22}+\rho_{33})$$
$$R_{03}=2(\rho_{11}+\rho_{33})-1\qquad R_{30}=2(\rho_{11}+\rho_{22})-1.$$
The trace distance quantum discord is invariant under  local
transformations and we have
\begin{equation*}
    D_T\left( \rho  \right) = D_T\left( {\tilde \rho } \right).
\end{equation*}
The minimization in the equation (\ref{eq1}) has been worked out for a generic two qubit-qubit $X$ state \cite{Ciccarello}.
As result, the trace distance quantum discord in the state $\rho$ Eq. (\ref{X}) takes the form
\begin{eqnarray}\label{eq3}
 D_{\rm T}(\rho)=\frac{1}{2}\sqrt{\frac{R_{11}^2
 R_{\rm max}^2-R_{22}^2R_{\rm min}^2}
 {R_{\rm max}^2-R_{\rm min}^2+R_{11}^2-R_{22}^2}}, \label{D}
\end{eqnarray}
where
$$R_{\rm min}^2=\min\{R_{11}^2,R_{33}^2\}\qquad {\rm with}\qquad R_{\rm max}^2={\rm \max} (R_{33}^2,R_{22}^2+R_{30}^2).$$
The Bell diagonal states constitute a specific instance of two-qubit $X$ states. In the Fano-Bloch
representation, they take the form
$$\rho^{\rm
BD}=\frac{1}{4}[I\otimes I+\vec{c}\cdot
(\vec{\sigma}\otimes\vec{\sigma})],$$
with $\vec{c}=\{c_1,c_2,c_3\}$
being a three-dimensional vector with elements satisfying $0
\leqslant |c_i| \leqslant 1$, and
$\vec{\sigma}=\{\sigma_1,\sigma_2,\sigma_3\}$ denotes the standard
Pauli matrices. The trace distance discord (\ref{D}) reduces to \cite{paula}
\begin{eqnarray}\label{eq4}
 D_{\rm T}(\rho^{\rm BD})={\rm int}\{|c_1|,|c_2|,|c_3|\},
\end{eqnarray}
that is the intermediate value for the absolute values of
the correlation factors $c_1$, $c_2$, and $c_3$.

\section{Local quantum uncertainty and trace discord of two qubits in independent reservoirs}

\subsection{The density matrix}
We consider a two-qubit system interacting with two independent reservoirs described by Ohmic-like spectral densities \cite{liu,Hao,Ferraro,Scala}. The whole system is described by the following Hamiltonian
\begin{equation}
    H = \sum_{j = 1}^2 {\left[ {\frac{{{v_j}}}{2}{\sigma _{j,3}} + \sum\limits_k {{w_{j,k}}b_{j,k}^\dagger {b_{j,k}} + \sum\limits_k {{\sigma _{j,3}}\left( {{g_{j,k}}b_{j,k}^\dagger  + g_{j,k}^*{b_{j,k}}} \right)} } } \right]}. \label{HH}
\end{equation}
The first term of this Hamiltonian describes the the two qubit $(j = 1, 2)$ with
${{v_j}}$ denoting the energy difference between the excited state  ${\left| 1 \right\rangle _j}$ and the ground state ${\left| 0 \right\rangle _j}$ and
     ${{\sigma _{j,3}}}$ is the third Pauli matrix.  The second term describes  two independent  reservoirs where $b_{j,k}^\dagger$ and ${{b_{j,k}}}$ are respectively  the bosonic creation and the annihilation operators. They satisfy the usual bosonic commutation relations and  ${{w_{j,k}}}$ denotes the frequency of the $k$-th mode of the reservoir coupled to  the qubit $j$. The last term in (\ref{HH}) expresses the interaction part between the reservoir and the qubit. The coupling strength between the qubit $j$ and the $k$-th mode is denoted by
      ${{g_{j,k}}}$.\\
To simplify our purpose,  we shall assume that
 the two qubits are initially prepared in the following Bell-diagonal states
  \begin{equation}
    {\rho _S}\left( 0 \right) = \frac{1}{4}\left( {I \otimes I + \sum\limits_{i = 1}^3 {{c_i}} {\sigma _i} \otimes {\sigma _i}} \right),
  \end{equation}
  where the correlation parameters satisfy the conditions $0 \le |{c_i}| \le 1$. Without loss of generality, we assume that  $|{c_1}|\geq|{c_2}$. The density matrix, in the computational basis $\left\{ {\left| {00} \right\rangle ,\left| {01} \right\rangle ,\left| {10} \right\rangle ,\left| {11} \right\rangle } \right\}$, rewrites as
   \begin{equation}
    \rho _S(0) = \frac{1}{4}
\left(
  \begin{array}{cccc}
    c^+_3 & 0 & 0 & c_- \\
    0 & c^-_3 & c_+ & 0 \\
    0 & c_+ & c^-_3 & 0 \\
    c_-& 0 & 0 & c^+_3 \\
  \end{array}
\right).
    \end{equation}
   where $c^+_3= 1+c_3$, $c^-_3= 1-c_3$, $c_+ = c_1 + c_2$ and $c_- = c_1 - c_2$.
The states of the reservoirs in thermal equilibrium at the temperature $T$ are given by
\begin{equation}
    \rho_{E_j} =  \frac{1}{Z_{E_j}}\exp \left( { - \beta \sum\limits_k {{w_{j,k}}b_{j,k}^\dagger {b_{j,k}}} } \right),
\end{equation}
where ${Z_{Ej}}$ is the partition function of the reservoir $j (j = 1, 2)$. Initially, the density matrix of the whole system reads as   $\rho \left( 0 \right) = {\rho _S}\left( 0 \right) \otimes {\rho _{{E_1}}} \otimes {\rho _{{E_2}}}$, and the evolved state is then given by
\begin{equation}
\rho (t) = \exp \left( { - iHt} \right)\rho \left( 0 \right)\exp \left( {iHt} \right).
\end{equation}
The density matrix of the two-qubit system is then obtained by tracing out the environment degrees of freedom. This gives
\begin{equation}
        \rho_S ( t) = {\rm Tr}_E\left[ {\exp \left( { - iHt} \right)\rho \left( 0 \right)\exp \left( {iHt} \right)} \right].
\end{equation}
The  matrix elements of ${\rho _S}\left( t \right)$ are given by
\begin{equation}
    \left\langle {ms} \right|{\rho _S}\left( t \right)\left| {nl} \right\rangle  = {\rm Tr}\left[ {{\pi _1}^{mn}\left( t \right){\pi _2}^{sl}\left( t \right)\rho \left( 0 \right)} \right],
\end{equation}
with ${\pi _j}^{ms}\left( t \right) = \exp \left( {iHt} \right){\pi _j}^{ms}\exp \left( { - iHt} \right)$ is the Heisenberg operator of
qubit $j$ and ${\pi _j}^{ms} = {\left| s \right\rangle _j}\left\langle m \right|$  where $s$ and $m$ take the values 0 and 1.\\

It is simple to see that the for $s=m$, the operators ${\pi_j}^{mm} = \frac{1}{2} ( 1 + (- 1)^m \sigma_3)$ commute with the
Hamiltonian $H$ and one has ${\pi _j}^{mm} (t) = {\pi_j}^{mm} (0)$. For different values of $s$ and $m$, the operators ${\pi _j}^{ms}$ are
given by  ${\pi _j}^{ms}(0) = \vert s \rangle_j \langle m \vert = \frac{1}{2} (\sigma_1 + i  (- 1)^m \sigma_2)$. Therefore
to determine the operators ${\pi _j}^{ms}(t)$ for $s \ne m$, one has to solve the following Heisenberg equations  for the bosonic modes
\begin{equation}
    i\frac{db_{j,k}(t)}{dt}  = {w_{j,k}}{b_{j,k}} ( t) + {g_{j,k}}{\sigma _{j,3}},
\end{equation}
and those associated with the two qubits are
\begin{equation}
    i\frac{d{\pi _j}^{ms}(t)}{dt} = {\left( { - 1} \right)^{m + 1}}\left[ { {v_j} + 2\sum\limits_k {\left( {{g_{j,k}}b_{j,k}^ \dagger \left( t \right) + g_{j,k}^*{b_{j,k}}\left( t \right)} \right)} } \right]{\pi _j}^{ms}\left( t \right).
\end{equation}
From the later equations, one gets
\begin{equation}
{\pi _j}^{01}(t) = \big({\pi _j}^{10}( t)\big)^{\dagger} = {\pi _j}^{01}\exp \left\{ {i{v_j}t - 2\sum\limits_k {\left( {a_{j,k}\left( t \right)b_{j,k}^\dagger  - a_{j,k}^*\left( t \right){b_{j,k}}} \right)} } \right\},
\end{equation}
where ${a_{j,k}} ( t ) = {{{g_{j,k}}\left( {1 - \exp \left( {i{w_{j,k}}t} \right)} \right)} \mathord{\left/
        {\vphantom {{2{g_{j,k}}\left( {1 - \exp \left( {i{w_{j,k}}t} \right)} \right)} {{w_{j,k}}}}} \right.
        \kern-\nulldelimiterspace} {{w_{j,k}}}}$.

In the Fano-Bloch representation, the matrix $\rho_S(t)$ can be expressed as follows
\begin{equation}\label{rhot}
    \rho_S(t) = \frac{1}{4}\bigg[ I \otimes I + T_{11} \sigma_1 \otimes \sigma_1 + {T_{12}}{\sigma_1} \otimes {\sigma _2} + {T_{21}}{\sigma _2} \otimes {\sigma _1} + {T_{22}}{\sigma _2} \otimes {\sigma _2} + {T_{33}}{\sigma_3} \otimes {\sigma _3} \bigg].
\end{equation}
The non-vanishing correlation parameters ${T_{\alpha \beta }}$, occurring in (\ref{rhot}), are given by:
\begin{equation}
    \begin{array}{l}
        {T_{11}} = \left[ {{c_1}\cos \left( {{v_1}t} \right)\cos \left( {{v_2}t} \right) + {c_2}\sin \left( {{v_1}t} \right)\sin \left( {{v_2}t} \right)} \right]{e^{ - \gamma \left( t \right)}}\\
        {T_{12}} = \left[ {{c_1}\cos \left( {{v_1}t} \right)\sin \left( {{v_2}t} \right) - {c_2}\sin \left( {{v_1}t} \right)\cos \left( {{v_2}t} \right)} \right]{e^{ - \gamma \left( t \right)}}\\
        {T_{21}} = \left[ {{c_1}\sin \left( {{v_1}t} \right)\cos \left( {{v_2}t} \right) - {c_2}\cos \left( {{v_1}t} \right)\sin \left( {{v_2}t} \right)} \right]{e^{ - \gamma \left( t \right)}}\\
        {T_{22}} = \left[ {{c_1}\sin \left( {{v_1}t} \right)\sin \left( {{v_2}t} \right) + {c_2}\cos \left( {{v_1}t} \right)\cos \left( {{v_2}t} \right)} \right]{e^{ - \gamma \left( t \right)}}\\
        {T_{33}} = {c_3}
    \end{array}
\end{equation}
where the time-dependent function $\gamma(t)$ is defined by:
\begin{equation}
    \gamma \left( t \right) = {\sum\limits_{j = 1}^2 {\sum\limits_k {4\left| {{g_{j,k}}} \right|} } ^2}{w_{j,k}}^{ - 2}\coth \left( {\frac{{\beta {w_{j,k}}}}{2}} \right)\left[ {1 - \cos \left( {{w_{j,k}}t} \right)} \right] \label{gamma}.
\end{equation}
In the continuum limit, the term ${\sum\limits_k {4\left| {{g_{j,k}}} \right|} ^2}$ is replaced by $\int {dw{J_j}\left( w \right)\delta \left( {{w_{j,k}} - w} \right)} $ and
to simplify our purpose we consider the situation where both reservoirs have the same spectral density ${J_1}\left( w \right) = {J_2}\left( w \right) = J\left( w \right)$.
In this limiting  case, ${\gamma \left( t \right)}$ can be expressed as:
\begin{equation}
    \gamma \left( t \right) = 2\int\limits_0^\infty  {dwJ\left( w \right){w^{ - 2}}\coth \left( {\frac{{\beta w}}{2}} \right)\left[ {1 - \cos \left( {wt} \right)} \right]}. \label{gama}
\end{equation}
The expression of the spectral density characterizing each reservoir is given by:
\begin{equation}
    J\left( w \right) = \lambda {\Omega ^{1 - s}}{w^s}{e^{ - {w \mathord{\left/
                    {\vphantom {w \Omega }} \right.
                    \kern-\nulldelimiterspace} \Omega }}},
\end{equation}
with $\Omega$ is the cutoff frequency and $\lambda$ is a dimensionless coupling constant between the system and the environment (the reservoirs).
 For $s = 1$, the reservoir is of ohmic type, for  $0 < s < 1$, the reservoir is  sub-ohmic and  for $s > 1$, the reservoir is called  super-ohmic.

\subsection{Dynamics of Local quantum uncertainty of two qubits in independent reservoirs}
In the computational basis, the density matrix (\ref{rhot}) takes the following form
{\footnotesize \begin{equation}\label{rhoS}
    \rho _S(t) = \frac{1}{4}
\left(
  \begin{array}{cccc}
    c^+_3 & 0 & 0 & c_-{e^{ - i\left( {{v_2} + {v_1}} \right)t - \gamma \left( t \right)}} \\
    0 & c^-_3 & c_+{e^{i\left( {{v_2} - {v_1}} \right)t - \gamma \left( t \right)}} & 0 \\
    0 & c_+{e^{-i\left( {{v_2} - {v_1}} \right)t - \gamma \left( t \right)}} & c^-_3 & 0 \\
    c_-{e^{  i\left( {{v_2} + {v_1}} \right)t - \gamma \left( t \right)}} & 0 & 0 & c^+_3 \\
  \end{array}
\right).
\end{equation}}
which is of two-qubit $X$ type. It follows that the Local quantum uncertainty and trace discord can be easily evaluated
using the results reported hereinabove. To obtain the explicit expression of local quantum uncertainty, one computes first the
elements of the matrix $W$. From (\ref{w11}), one gats

{\normalsize  \begin{align}
        w_{11}&= \frac{1}{2} \sqrt{\bigg( c^+_3 + \sqrt{(c^+_3)^2 - (c_-)^2 e^{-2\gamma(t)}}\bigg)\bigg( c^-_3 + \sqrt{(c^-_3)^2 - (c_+)^2 e^{-2\gamma(t)}}\bigg)}\notag \\
        &+\frac{1}{2} \frac{(c_-)^2 e^{-2\gamma(t)}\cos(2v_1t)}{\sqrt{\bigg( c^+_3 + \sqrt{(c^+_3)^2 - (c_-)^2 e^{-2\gamma(t)}}\bigg)\bigg( c^-_3 + \sqrt{(c^-_3)^2 - (c_+)^2 e^{-2\gamma(t)}}\bigg)}}.
\end{align}   }
Using the expression (\ref{w22}), one obtains
{\normalsize  \begin{align}
        {w_{22}} &= \frac{1}{2}\left( {\sqrt {\left( { {c_3^+} + \sqrt {{{\left( {c_3^+} \right)}^2} - {{\left( {c_-} \right)}^2}{e^{ - 2\gamma \left( t \right)}}} } \right)\left( {c_3^- + \sqrt {{{\left( {c_3^-} \right)}^2} - {{\left( {c_+} \right)}^2}{e^{ - 2\gamma \left( t \right)}}} } \right)} } \right) \notag \\
        &- \frac{1}{2}\left( {\frac{{{{\left( {c_-} \right)}^2}{e^{ - 2\gamma \left( t \right)}}\cos \left( {2{v_1}t} \right)}}{{\sqrt {\left( {c_3^+ + \sqrt {{{\left( {c_3^+} \right)}^2} - {{ {c_-} }^2}{e^{ - 2\gamma \left( t \right)}}} } \right)\left( {c_3^- + \sqrt {{{\left( {c_3^-} \right)}^2} - {{\left( {c_+} \right)}^2}{e^{ - 2\gamma \left( t \right)}}} } \right)} }}} \right)
\end{align}   }
The equation (\ref{w33}) gives
{\normalsize  \begin{align}
        {w_{33}} &= \frac{1}{4}\left( {2 + \sqrt {{{\left( {c_3^+} \right)}^2} - {{\left( {c_-} \right)}^2}{e^{ - 2\gamma \left( t \right)}}}  + \sqrt {{{\left( {c_3^-} \right)}^2} - {{\left( {c_+} \right)}^2}{e^{ - 2\gamma \left( t \right)}}} } \right) \notag \\
        &- \frac{{{{\left( {c_-} \right)}^2}{e^{ - 2\gamma \left( t \right)}}}}{{4\left( {c_3^+ + \sqrt {{{\left( {c_3^+} \right)}^2} - {{\left( {c_-} \right)}^2}{e^{ - 2\gamma \left( t \right)}}} } \right)}} - \frac{{{{\left( {c_+} \right)}^2}{e^{ - 2\gamma \left( t \right)}}}}{{4\left( {c_3^-+ \sqrt {{{\left( {c_3^-} \right)}^2} - {{\left( {c_+} \right)}^2}{e^{ - 2\gamma \left( t \right)}}} } \right)}}
\end{align}}
The non vanishing  off-diagonal elements (\ref{w12}) write
\begin{equation}
    \normalsize { { {w_{12}} = {w_{21}} = \frac{{{c_-c_+} {e^{ - 2\gamma \left( t \right)}}\cos \left( {{v_1}t} \right)\sin \left( {{v_1}t} \right)}}{{\sqrt {\left( {c_3^+ + \sqrt {{{\left( {c_3^+} \right)}^2} - {{\left( {c_-} \right)}^2}{e^{ - 2\gamma \left( t \right)}}} } \right)\left( {c_3^- + \sqrt {{{\left( {c_3^- } \right)}^2} - {{\left( {c_+} \right)}^2}{e^{ - 2\gamma \left( t \right)}}} } \right)} }}}}.
\end{equation}
 The eigenvalues of the matrix $W$ are
\begin{equation}
    {\lambda _1} = \frac{1}{2}\left[ {\left( {{w_{11}} + {w_{22}}} \right) + \sqrt {{{\left( {{w_{11}} + {w_{22}}} \right)}^2} - 4\left( {{w_{11}}{w_{22}} - {w_{21}}{w_{12}}} \right)} } \right],
\end{equation}
\begin{equation}
    {\lambda _2} = \frac{1}{2}\left[ {\left( {{w_{11}} + {w_{22}}} \right) - \sqrt {{{\left( {{w_{11}} + {w_{22}}} \right)}^2} - 4\left( {{w_{11}}{w_{22}} - {w_{21}}{w_{12}}} \right)} } \right],
\end{equation}
\begin{equation}
    {\lambda _3} = w_{33}
\end{equation}
We have assumed that $\vert c_2 \vert \leq \vert c_1 \vert$. Thus,  one has $\lambda _2 \leq \lambda_1$ and subsequently the local quantum uncertainty is given in term of
\begin{equation}
    {\lambda _{\max }} = {\rm max} (\lambda_1,\lambda_3).
\end{equation}
Therefore, one should consider two cases:  $\lambda_1 \leq \lambda_3$ and $\lambda_3 \leq \lambda_1$. Hence, when  $\lambda_{\rm max}=\lambda_{1}$, the local quantum
uncertainty is given by the following expression
\begin{align}
{U_A}\left( \rho_S(t) \right) &=\frac{1}{2}\left[ {2 - \sqrt {\left( {c_3^+ + \sqrt {{{\left( {c_3^+} \right)}^2} - {{\left( {c_-} \right)}^2}{e^{ - 2\gamma \left( t \right)}}} } \right)\left( {c_3^- + \sqrt {{{\left( {c_3^-} \right)}^2} - {{\left( {c_+} \right)}^2}{e^{ - 2\gamma \left( t \right)}}} } \right)} } \right] \notag\\
& - \frac{{ {c_-c_+} {e^{ - 2\gamma \left( t \right)}}}}{{2\sqrt {\left( {c_3^+ + \sqrt {{{\left( {c_3^+} \right)}^2} - {{\left( {c_-} \right)}^2}{e^{ - 2\gamma \left( t \right)}}} } \right)\left( {c_3^- + \sqrt {{{\left( {c_3^-} \right)}^2} - {{\left( {c_+} \right)}^2}{e^{ - 2\gamma \left( t \right)}}} } \right)} }},
\end{align}
and when $\lambda_{max}=\lambda_{3}$, one gets
\begin{align}
{U_A}\left( \rho_S(t) \right)& = \frac{1}{4}\left[ {2 - \sqrt {{{\left( {c_3^+} \right)}^2} - {{\left( {c_-} \right)}^2}{e^{ - 2\gamma \left( t \right)}}}  - \sqrt {{{\left( {c_3^-} \right)}^2} - {{\left( {c_+} \right)}^2}{e^{ - 2\gamma \left( t \right)}}} } \right] \notag\\
&+ \frac{{{{\left( {c_-} \right)}^2}{e^{ - 2\gamma \left( t \right)}}}}{{4\left[ {c_3^+ + \sqrt {{{\left( {c_3^+} \right)}^2} - {{\left( {c_-} \right)}^2}{e^{ - 2\gamma \left( t \right)}}} } \right]}} +
 \frac{{{{\left( {c_+} \right)}^2}{e^{ - 2\gamma \left( t \right)}}}}{{4\left[ {c_3^- + \sqrt {{{\left( {c_3^-} \right)}^2} - {{\left( {c_+} \right)}^2}{e^{ - 2\gamma \left( t \right)}}} } \right]}}.
\end{align}

\subsection{Dynamics of trace quantum discord of two qubits in independent reservoirs}
The non vanishing matrix correlations, in the Fano-Bloch representation, of the density matrix $\rho_S(t)$ (\ref{rhoS}) are gien by
$$R_{11}=c_{1}e^{-\gamma\left(t \right) } \qquad R_{22}=c_{2}e^{-\gamma\left(t \right) } \qquad  R_{33}=c_{3}$$
The trace quantum discord (\ref{D}) takes the form
\begin{equation}
D_{T}\left(\rho_S(t) \right)=\frac{e^{-\gamma\left( t\right) }}{2}\sqrt {\frac{c_{1}^{2}\max\{c_{3}^{2},c_{2}^{2} e^{-2\gamma\left( t\right) }\}-c_{2}^{2}\min\{c_{1}^{2} e^{-2\gamma\left( t\right) },c_{3}^{2}\}}{\max\{c_{3}^{2},c_{2}^{2} e^{-2\gamma\left( t\right) }\}-\min\{c_{1}^{2} e^{-2\gamma\left( t\right) },c_{3}^{2}\}+\left(c_{1}^{2}-c_{2}^{2} \right)e^{-2\gamma\left( t\right) } }}.
\end{equation}
For $|c_{3}|\geq|c_{2}|e^{-\gamma\left( t\right) }$ and $|c_{3}|\geq|c_{1}|e^{-\gamma\left( t\right) }$, the trace distance discord is given by:
\begin{equation}
D_{T}\left( \rho_S(t)\right)= \frac{1}{2} |c_{1}|e^{-\gamma\left( t\right) }.
\end{equation}
For $|c_{3}|\geq|c_{2}|e^{-\gamma\left( t\right) }$ and $|c_{3}|\leq|c_{1}|e^{-\gamma\left( t\right) }$ one gets:
\begin{equation}
D_{T}\left( \rho_S(t)\right)=\frac{1}{2} |c_{3}|.
\end{equation}
In this situation, it is remarkable that the quantum correlations are unaffected by the noisy
environment and the geometric discord exhibits a freezing behavior.
For  $|c_{3}|\leq|c_{2}|e^{-\gamma\left( t\right) }$ and $|c_{3}|\geq|c_{1}|e^{-\gamma\left( t\right) }$ one hase
 $$D_{T}\left( \rho_S(t) \right)=0,$$
 reflecting the absence of quantum correlations.
For $|c_{3}|\leq|c_{2}|e^{-\gamma\left( t\right) }$ and $|c_{3}|\leq|c_{1}|e^{-\gamma\left( t\right) }$, the trace distance discord is simply given by
\begin{equation}
D_{T}\left( \rho_S(t)\right)=    \frac{1}{2} |c_{2}|e^{-\gamma\left( t\right) }.
\end{equation}
\begin{figure}[H]
    \centering
    \begin{minipage}[t]{3in}
        \centering
        \includegraphics[width=3.1in]{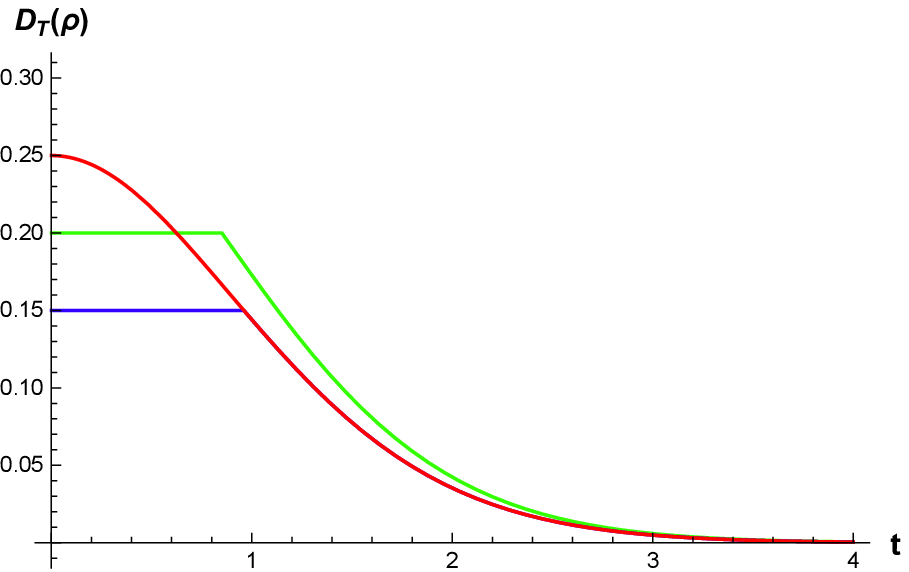}
    \end{minipage}
    \begin{minipage}[t]{3in}
        \centering
        \includegraphics[width=3.1in]{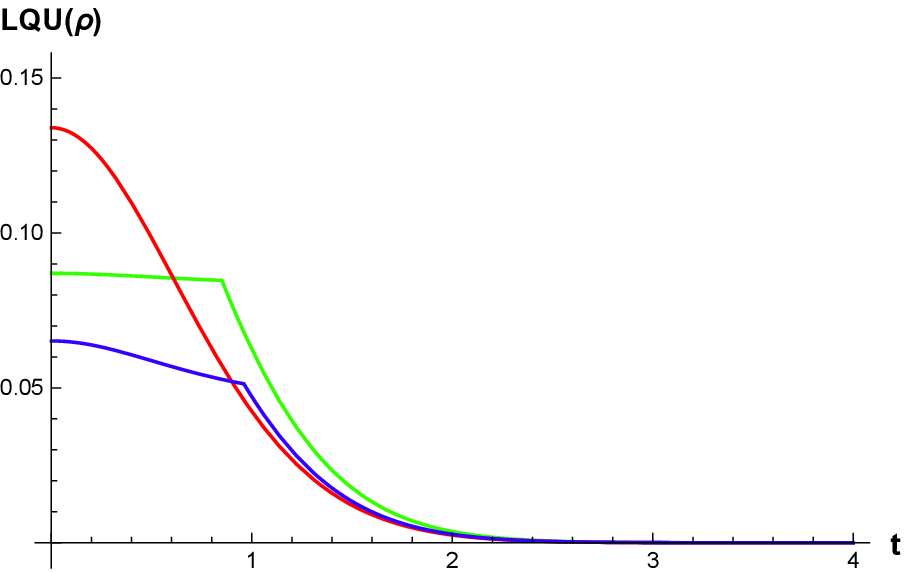}
    \end{minipage}

    {\bf Fig 1:} {\sf The local quantum uncertainty and trace distance discord for the sub-ohmic reservoirs with $s=0.5$ , $\lambda=0.1$, $\varOmega \beta=1$. $c_{1}=0.6$, $c_{2}=-0.3$, $c_{3}=0.4$ (green line). $c_{1}=-0.5$, $c_{2}=0$, $c_{3}=0.3$ (blue line).$c_{1}=0.5$, $c_{2}=-0.3$, $c_{3}=0.6$ (red line).}
\end{figure}
\begin{figure}[H]
    \centering
    \begin{minipage}[t]{3in}
        \centering
        \includegraphics[width=3.1in]{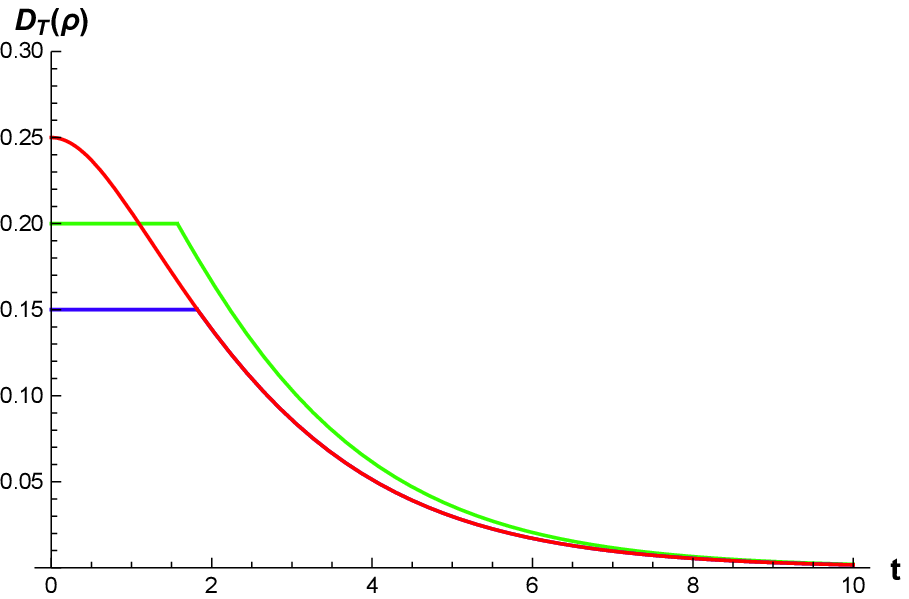}
    \end{minipage}
    \begin{minipage}[t]{3in}
        \centering
        \includegraphics[width=3.1in]{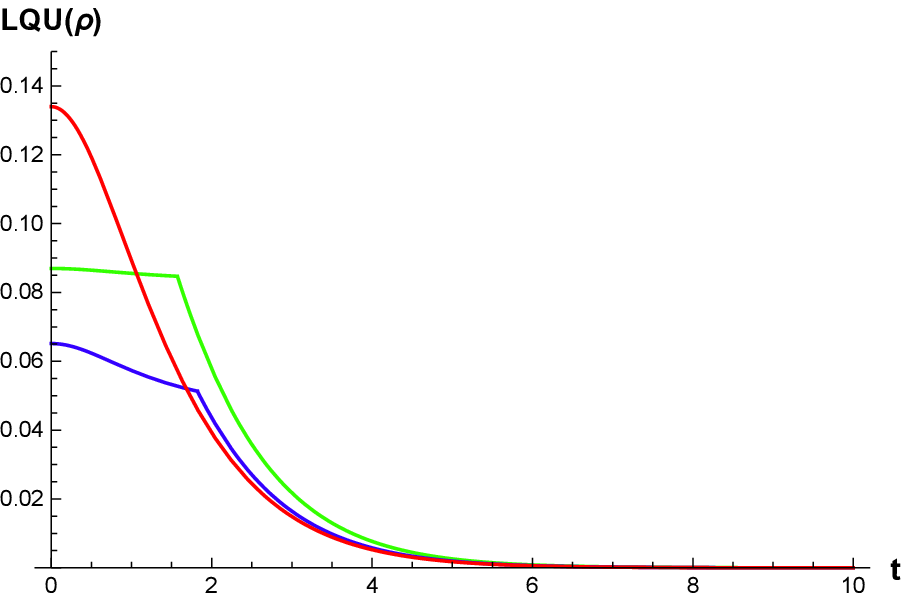}
    \end{minipage}

    {\bf Fig 2:} {\sf The local quantum uncertainty and trace distance discord for the ohmic reservoirs with $s=1$ , $\lambda=0.1$, $\varOmega \beta=1$. $c_{1}=0.6$, $c_{2}=-0.3$, $c_{3}=0.4$ (green line). $c_{1}=-0.5$, $c_{2}=0$, $c_{3}=0.3$ (blue line).$c_{1}=0.5$, $c_{2}=-0.3$, $c_{3}=0.6$ (red line).}
\end{figure}
\begin{figure}[H]
    \centering
    \begin{minipage}[t]{3in}
        \centering
        \includegraphics[width=3.1in]{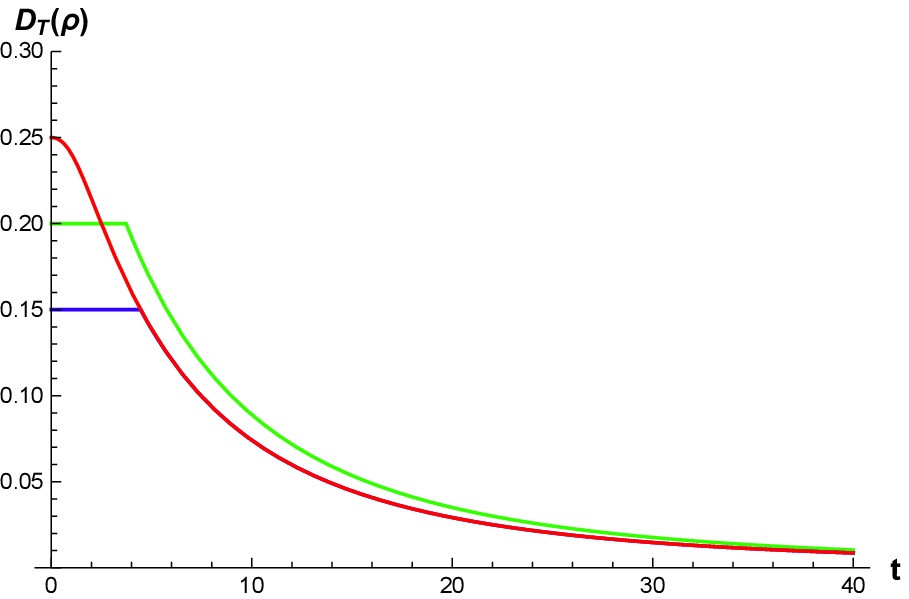}
    \end{minipage}
    \begin{minipage}[t]{3in}
        \centering
        \includegraphics[width=3.1in]{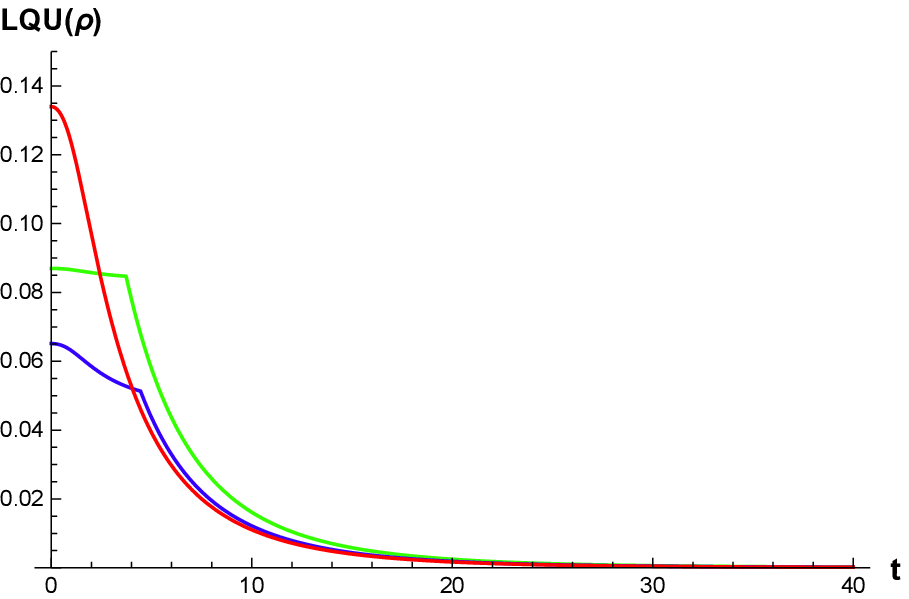}
    \end{minipage}

    {\bf Fig 3:} {\sf The local quantum uncertainty and trace distance discord for the super-ohmic with $s=1.5$ , $\lambda=0.2$, $T=0$. $c_{1}=0.6$, $c_{2}=-0.3$, $c_{3}=0.4$ (green line). $c_{1}=-0.5$, $c_{2}=0$, $c_{3}=0.3$ (blue line).$c_{1}=0.5$, $c_{2}=-0.3$, $c_{3}=0.6$ (red line).}
\end{figure}
We begin by analyzing the behavior of local quantum uncertainty in three particular Bell states coupled to environment of ohmic, sub-ohmic and   super-ohmic type. As depicted in the figures 1, 2 and 3,
the local quantum uncertainty may exhibit a sudden change behavior for the states with $(c_{1}=0.6, c_{2}=-0.3, c_{3}=0.4)$ and $(c_{1}=-0.5, c_{2}=0, c_{3}=0.3)$. This behavior is not observable for
the state with $(c_{1}=0.5, c_{2}=-0.3, c_{3}=0.6)$ for which the local quantum uncertainty decays monotonically. This decrease becomes more pronounced when passing from sub-ohmic to super-ohmic regime. This is essentially due to the  strong nature of  the environment effect in the sub-ohmic regime. It must be noticed also that for the states with $(c_{1}=0.6, c_{2}=-0.3, c_{3}=0.4)$ and $(c_{1}=-0.5, c_{2}=0, c_{3}=0.3)$,
the local quantum uncertainty varies almost linearly before the sudden change point. The interval of time, in which this variation is linear,  depends
on the nature of the system-environment coupling. It is  larger in the ohmic regime in comparison with the others regimes. Let us now analyze the behavior of quantum correlations measured by trace norm. We notice first that the local quantum uncertainty and trace discord exhibit similar behavior
for the considered states, except the freezing behavior which occurs before the sudden change of trace norm. We obtain a large and controllable freezing interval for states with $(c_{1}=0.6, c_{2}=-0.3, c_{3}=0.4)$ evolving in Markovian environment of super-ohmic type. This interval is reduced by the decoherence effects in the case of ohmic and sub-ohmic regimes. This freezing phenomenon exhibited by trace norm  reflects that the quantum correlations in a given quantum state are not affected by the noise generated by its surrounding environment. This is surprising and questionable phenomenon does not occur when one consider local quantum uncertainty as quantifier of quantum discord. Indeed, as we have discussed above, in the interval where the trace discord is constant, the local quantum uncertainty decreases linearly until the point when the sudden change of the behavior happens.

\section{Quantum correlations dynamics of two 2-level atoms interacting with an electromagnetic field}
In this section, we consider two identical atoms with ground states $\left| {{g_i}} \right\rangle $ and excited states $\left| {{e_i}} \right\rangle$ ($i = 1, 2$) which are  coupled  with a quantized electromagnetic field \cite{Agarwal,Tanas,Ficek,Auyuanet}. In
the  rotating-wave approximation, the Hamiltonian can write
\begin{equation}
    \hat H = \hbar {\omega _0}{S^z} + \sum\limits_{\vec ks} {{\omega _k}\hat a_{\vec ks}^\dag {{\hat a}_{\vec ks}}}  - i\hbar \sum\limits_{\vec ks} {\left[ {\vec \mu .{{\vec g}_{\vec ks}}{S^ + }{{\hat a}_{\vec ks}} - H.c} \right]},
\end{equation}
where ${{{\hat a}_{\vec ks}}}$ and ${\hat a_{\vec ks}^\dag }$ are respectively the annihilation and creation operators corresponding to the field mode ${\vec ks}$ , which has wave
vector $\vec{k}$, frequency $\omega_k$ and the index of polarization $s$. The coupling factor is given by
$${{\vec g}_{\vec ks}}\left( {{{\vec r}_i}} \right) = {\left( {\frac{{{\omega _k}}}{{2{\varepsilon _0}\hbar V}}} \right)^{\frac{1}{2}}}{{\vec e}_{_{\vec ks}}}{e^{i\vec k.{{\vec r}_i}}}$$
where $V$ denotes the quantization volume and $e_{ks}$ is  the electric field
polarization vector. This factor  represents the  mode function of the three dimensional
field, evaluated at the position $\vec{r}_i$ of the ith atom. The quantity $\vec \mu$ is the transition dipole moment and ${\omega _0}$ is the transition frequency.
The operators ${S^ \pm }$ and ${S^z}$ are the collective spin operators defined by ${S^ \pm } = \sum\limits_i {S_i^ \pm } $ and ${S^z} = \sum\limits_i {S_i^z} $ with $S_i^ +  = \left| {{e_i}} \right\rangle \left\langle {{g_i}} \right|$, $S_i^ -  = \left| {{g_i}} \right\rangle \left\langle {{e_i}} \right|$ and $S_i^z = \left| {{e_i}} \right\rangle \left\langle {{e_i}} \right| - \left| {{g_i}} \right\rangle \left\langle {{g_i}} \right|$.\\
To study the dynamics of this system, we shall employ the following  master equation \cite{Tanas}:
\begin{equation}\label{master-eq}
    \frac{{\partial \rho \left( \tau  \right)}}{{\partial t}} =  - i{\omega _0}\sum\limits_{i = 1}^2 {\left[ {S_i^z,\rho } \right]}  - i{\Omega _{12}}\sum\limits_{i \ne j}^2 {\left[ {S_i^ + S_j^ - ,\rho } \right]}  - \frac{1}{2}\sum\limits_{i,j = 1}^2 {{\Gamma _{ij}}\left( {\rho S_i^ + S_j^ -  + S_i^ + S_j^ - \rho  - 2S_j^ - \rho S_i^ + } \right)}
\end{equation}

where ${\Gamma _{ii}}\equiv \Gamma$ $(i = 1, 2)$ is  the spontaneous emission rate induced by the
direct coupling of the atom with the radiation field,  ${\Gamma _{12}} = {\Gamma _{21}}$ denotes the collective damping and $\Omega_{12}$ is the dipole-dipole interaction potential. They are  given by \cite{Tanas,Ficek,Auyuanet}:
\begin{equation}
    {\Gamma _{ij}} = \frac{3}{2}\Gamma \left\{ {\left[ {1 - {{\left( {\vec \mu .{{\vec r}_{ij}}} \right)}^2}} \right]\frac{{\sin \left( {{k_0}{r_{ij}}} \right)}}{{{k_0}{r_{ij}}}} + \left[ {1 - 3{{\left( {\vec \mu .{{\vec r}_{ij}}} \right)}^2}} \right]\left[ {\frac{{\cos \left( {{k_0}{r_{ij}}} \right)}}{{{{\left( {{k_0}{r_{ij}}} \right)}^2}}} - \frac{{\sin \left( {{k_0}{r_{ij}}} \right)}}{{{{\left( {{k_0}{r_{ij}}} \right)}^3}}}} \right]} \right\},
\end{equation}
and
\begin{equation}
    {\Omega _{ij}} = \frac{3}{4}\Gamma \left\{ { - \left[ {1 - {{\left( {\vec \mu .{{\vec r}_{ij}}} \right)}^2}} \right]\frac{{\cos \left( {{k_0}{r_{ij}}} \right)}}{{{k_0}{r_{ij}}}} + \left[ {1 - 3{{\left( {\vec \mu .{{\vec r}_{ij}}} \right)}^2}} \right]\left[ {\frac{{\sin \left( {{k_0}{r_{ij}}} \right)}}{{{{\left( {{k_0}{r_{ij}}} \right)}^2}}} + \frac{{\cos \left( {{k_0}{r_{ij}}} \right)}}{{{{\left( {{k_0}{r_{ij}}} \right)}^3}}}} \right]} \right\}.
\end{equation}
where ${r_{ij}} = \left| {{r_j} - {r_i}} \right|$ is the distance between the atoms, and ${k_0} = \frac{{{\omega _0}}}{c}$.
We consider the situation where the two qubits are initially
prepared in their excited states $\vert e_1 , e_2 \rangle$. Using the master equation (\ref{master-eq}), the evolved two-qubit state writes in the computational basis as
\begin{equation}\label{rhotau}
    \rho \left( \tau  \right) = \left( {\begin{array}{*{20}{c}}
            {a\left( \tau  \right)}&0&0&0 \\
            0&{b\left( \tau  \right)}&{c\left( \tau  \right)}&0 \\
            0&{c\left( \tau  \right)}&{b\left( \tau  \right)}&0 \\
            0&0&0&{1 - a\left( \tau  \right) - 2b\left( \tau  \right)}
    \end{array}} \right).
\end{equation}
where
\begin{equation} \label{adetau}
    a\left( \tau  \right) = {e^{ - 2\tau }},
\end{equation}
\begin{equation}\label{bdetau}
    b\left( \tau  \right) = \frac{{{e^{ - \tau }}}}{{\left( {1 - \gamma } \right)}}\left[ {\left( {1 + {\gamma ^2}} \right)\left( {\cosh\left( {\gamma \tau } \right) - {e^{ - \tau }}} \right) - 2\gamma \sinh\left( {\gamma \tau } \right)} \right],
\end{equation}
\begin{equation}\label{cdetau}
    c\left( \tau  \right) = \frac{{{e^{ - \tau }}}}{{\left( {1 - \gamma } \right)}}\left[ {2\gamma \left( {\cosh\left( {\gamma \tau } \right) - {e^{ - \tau }}} \right) - \left( {1 + {\gamma ^2}} \right)\sinh\left( {\gamma \tau } \right)} \right],
\end{equation}
with  $\tau = \Gamma t$ and $\gamma  = {{{\Gamma _{12}}} \mathord{\left/
        {\vphantom {{{\Gamma _{12}}} \Gamma }} \right.
        \kern-\nulldelimiterspace} \Gamma }$. \\
The density matrix $\rho \left( \tau  \right)$ rewrites, in Fano-Bloch representation,  as
\begin{equation}
    \rho \left( \tau  \right) = \frac{1}{4}\sum\limits_{\alpha ,\beta } {{T_{\alpha \beta }}} {\sigma _\alpha } \otimes {\sigma _\beta }
\end{equation}
where the nonvanishing correlation matrix elements are given by
\begin{equation}\label{T11T22}
{T_{11}} = {T_{22}} = 2c\left( \tau  \right) \hspace{1cm} {T_{33}} = 1 - 4b\left( \tau  \right) \hspace{1cm} {T_{03}} = {T_{30}} = 2\left( {a\left( \tau  \right) + b\left( \tau  \right)} \right) - 1,
\end{equation}
in term of the time dependent functions $a(\tau)$, $b(\tau)$ and $c(\tau)$ given respectively by (\ref{adetau}), (\ref{bdetau}) and (\ref{cdetau}).

\subsection{Dynamics of Local quantum uncertainty}
To determine the local quantum uncertainty for the density matrix $\rho(\tau)$, one employs the results obtained in section 2. Thus,  the elements of the matrix $W$ (\ref{w-elements}) are given by
\begin{equation}
    {w_{11}} = {w_{22}} = \left( {\sqrt {b\left( \tau  \right) + c\left( \tau  \right)}  + \sqrt {b\left( \tau  \right) - c\left( \tau  \right)} } \right)\left( {\sqrt {a\left( \tau  \right)}  + \sqrt {1 - a\left( \tau  \right) - 2b\left( \tau  \right)} } \right),
\end{equation}
\begin{equation}
    {w_{33}} = 1 - 2b\left( \tau  \right) + 2\sqrt {b{{\left( \tau  \right)}^2} - c{{\left( \tau  \right)}^2}},
\end{equation}
which can be rewritten also as
\begin{align*}
    {w_{11}} = {w_{22}}& = \frac{{{e^{ - \tau }}}}{{\left( {1 - {\gamma ^2}} \right)}}\left[ {\sqrt {\left( {1 - {\gamma ^2}} \right){e^{ - \tau }}}  + \sqrt {4\gamma \sinh\left( {\gamma \tau } \right) + 2\left( {1 - {\gamma ^2}} \right)\sinh\left( \tau  \right) - 2\left( {1 + {\gamma ^2}} \right)\left( {\cosh\left( {\gamma \tau } \right) - {e^{ - \tau }}} \right)} } \right] \\
    & \hspace{2.5cm} \left[ {\left( {1 + \gamma } \right)\sqrt {{e^{ - \gamma \tau }} - {e^{ - \tau }}}  + \left( {1 - \gamma } \right)\sqrt {{e^{\gamma \tau }} - {e^{ - \tau }}} } \right],
\end{align*}
\begin{equation*}
    {w_{33}} = 1 + \frac{{2{e^{ - \tau }}}}{{\left( {1 - {\gamma ^2}} \right)}}\left[ {2\gamma \sinh\left( {\gamma \tau } \right) - \left( {1 + {\gamma ^2}} \right)\left( {\cosh\left( {\gamma \tau } \right) - {e^{ - \tau }}} \right) + \sqrt {1 + {e^{ - 2\tau }} - 2{e^{ - \tau }}\cosh\left( {\gamma \tau } \right)} } \right].
\end{equation*}
In the situation where the wo atoms are very close ( ${r_{12}} \to 0$), we have ${\Gamma _{12}} \to \Gamma $. This  means that $\gamma  \to 1$ and $b\left( \tau  \right) = c\left( \tau  \right) = {e^{ - 2\tau }}\tau $.
In this case,  the elements of the matrix $W$ are simply given by
\begin{equation}
    {w_{11}} = {w_{22}} = \sqrt {2b\left( \tau  \right)a\left( \tau  \right)}  + \sqrt {2b\left( \tau  \right)\left( {1 - a\left( \tau  \right) - 2b\left( \tau  \right)} \right)}
\end{equation}
\begin{equation}
    {w_{33}} = 1 - 2b\left( \tau  \right)
\end{equation}
which rewrite as
$${w_{11}} = {w_{22}} = \sqrt {2{e^{ - 2\tau }}\tau } \left[ {\sqrt {{e^{ - 2\tau }}}  + \sqrt {1 - {e^{ - 2\tau }}\left( {1 + 2\tau } \right)} } \right]\hspace{1cm} {\rm and} \hspace{1cm} {w_{33}} = 1 - 2{e^{ - 2\tau }}\tau .$$

\subsection{Dynamics of geometric quantum discord}
To determine the analytic expression of the trace distance discord
of this system from the equation (\ref{eq3}), it is necessary to calculate the expressions of
$T_{max}^{2}$ and $T_{min}^{2}$ by comparing
$T_{33}^{2}$ with $T_{22}^{2}+T_{30}^{2}$ and $T_{11}^{2}$ with
$T_{33}^{2}$. It is simple to check that the difference $T_{33}^{2}-T_{22}^{2}-T_{30}^{2}$ is negative and $T_{max}^{2}= T_{22}^{2} + T_{30}^{2} $ . Therefore, to get
the trace discord, one has to treat separately the cases  $|T_{33}|\geq |T_{11}|$ and $|T_{11}|\geq |T_{33}|$. Using the equation (\ref{eq3}) and noticing that $T_{11} = T_{22}$ (see Eq.(\ref{T11T22})),
one verifies that  the trace discord,  in both cases, writes
\begin{equation}
D_{T}\left( \rho(\tau)\right)= \frac{1}{2} \vert T_{11}\vert = \vert c(\tau)\vert,
\end{equation}
where $c(\tau)$ is given  by (\ref{cdetau}).

\begin{figure}[H]
    \centering
    \begin{minipage}[t]{3in}
        \centering
        \includegraphics[width=3.1in]{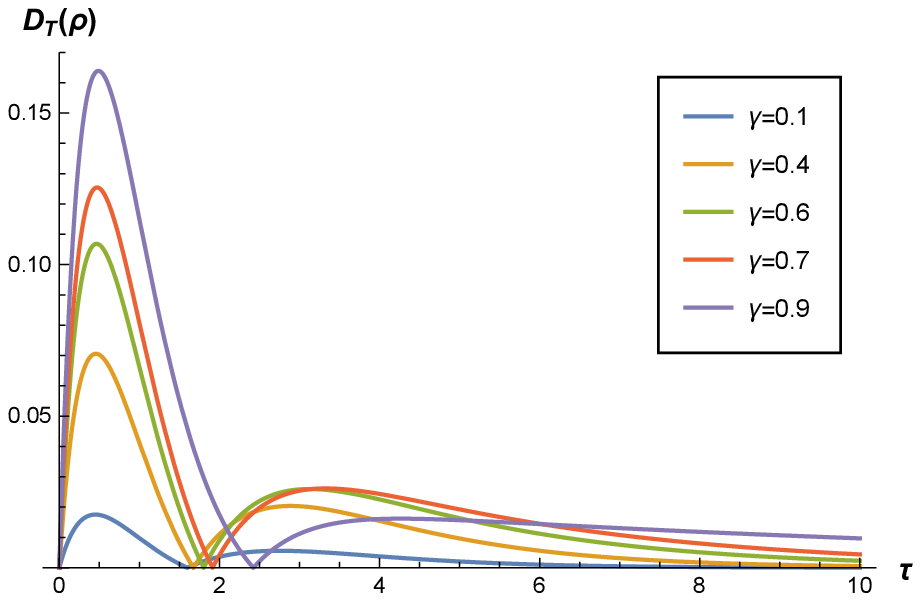}
    \end{minipage}
    \begin{minipage}[t]{3in}
        \centering
        \includegraphics[width=3.1in]{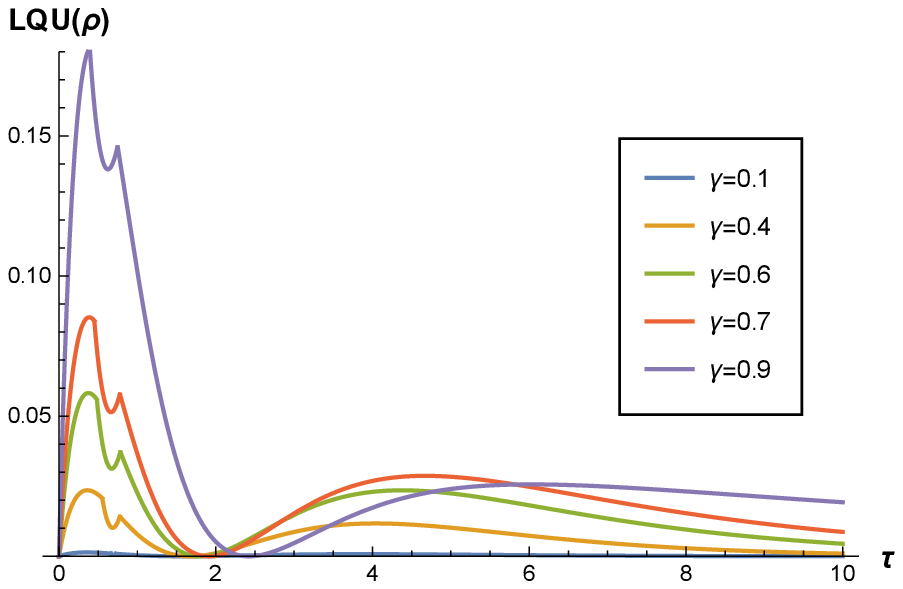}
    \end{minipage}\\
    {\bf Fig 4:} {\sf The trace distance discord $D_{T}\left(\rho\right)$ and local quantum uncertainty versus the parameter $\tau$ for the different values of $\gamma$.}
    \label{Fig.1}
\end{figure}
\begin{figure}[H]
    \centerline{\includegraphics[width=9cm]{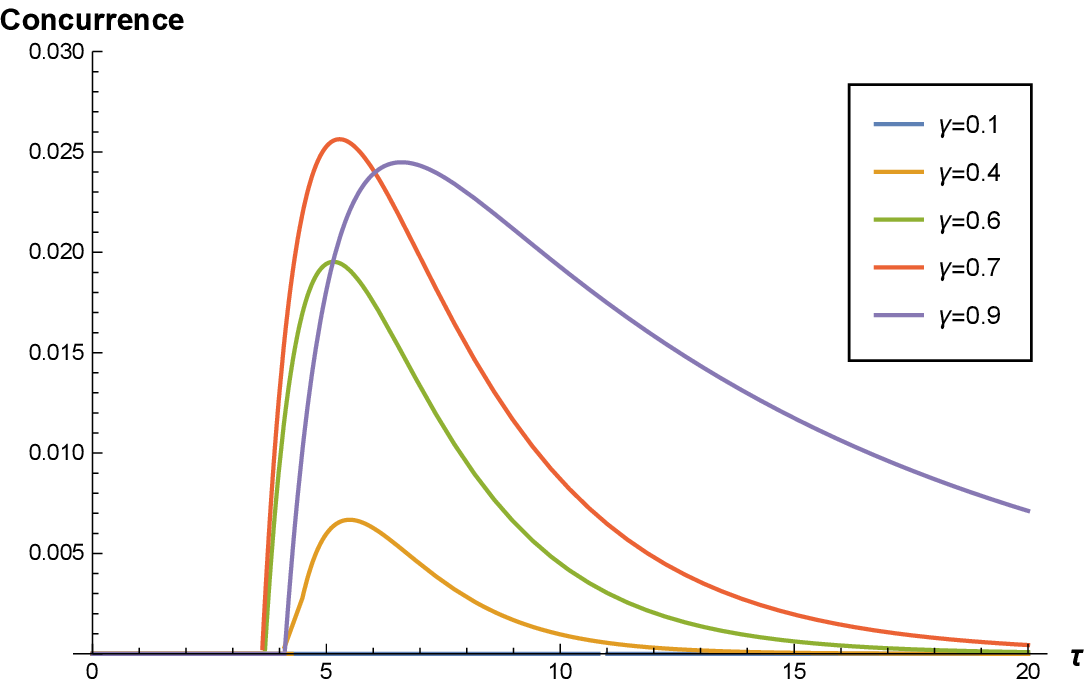}}
    {\bf Fig 5:} {\sf The variation of the Concurrence for the different values of $\gamma$.}
\end{figure}

\subsection{Dynamics of Concurrence}
The two atoms are initially prepared in the separable state   $\vert e_1 , e_2 \rangle$. Besides the local quantum uncertainty and trace quantum discord, we shall
also consider the dynamics of the concurrence. This measure, introduced by Wootters \cite{wigner,luo} is given by
\begin{equation}\label{defC}
C = \max \left\{ {0,{\lambda _1} - {\lambda _2} - {\lambda _3} - {\lambda _4}} \right\},
\end{equation}
where ${\lambda _1} \ge {\lambda _2} \ge {\lambda _3} \ge {\lambda _4}$ are the square roots of the eigenvalues of the matrix $\rho \tilde \rho $. The matrix $\tilde \rho $  that is obtained  by  the  "spin-flipped" operation from
the density matrix $\rho$ as
$\tilde \rho  = \left( {{\sigma _y} \otimes {\sigma _y}} \right){\rho ^*}\left( {{\sigma _y} \otimes {\sigma _y}} \right)$.
For the density matrix (\ref{rhotau}), the corresponding eigenvalues are
\begin{equation}
\{ \lambda_1= \lambda_2, \lambda_3, \lambda_4 \} = \left\{ {\sqrt {a\left( \tau  \right)\left( {1 - a\left( \tau  \right) - 2b\left( \tau  \right)} \right)} ,\left| {b\left( \tau  \right) - c\left( \tau  \right)} \right|,\left| {b\left( \tau  \right) + c\left( \tau  \right)} \right|} \right\},
\end{equation}
and the concurrence (\ref{defC}) writes
\begin{equation}
C\left(\rho\right)= \left\{ \begin{array}{l}
    \max \left\{ {0, - 2b\left( \tau  \right)} \right\} \hspace{4.5cm} {\rm if} \hspace{0.5cm}{\lambda _1} = \sqrt {a\left( \tau  \right)\left( {1 - a\left( \tau  \right) - 2b\left( \tau  \right)} \right)}, \\
    \max \left\{ {0, - 2\left( {c\left( \tau  \right) + \sqrt {a\left( \tau  \right)\left( {1 - a\left( \tau  \right) - 2b\left( \tau  \right)} \right)} } \right)} \right\} \hspace{1.2cm} {\rm if} \hspace{0.5cm}{\lambda _1}  = \left| {b\left( \tau  \right) - c\left( \tau  \right)} \right|,\\
    \max \left\{ {0,2\left( {c\left( \tau  \right) - \sqrt {a\left( \tau  \right)\left( {1 - a\left( \tau  \right) - 2b\left( \tau  \right)} \right)} } \right)} \right\}\hspace{1.5cm} {\rm if}\hspace{0.5cm}{\lambda _1}  = \left| {b\left( \tau  \right) + c\left( \tau  \right)} \right|,
    \end{array} \right..
\end{equation}
The dynamics of local quantum uncertainty and trace quantum discord are plotted in the Figure 4.   As depicted in Fig. 4, the behavior of both the local quantum uncertainty and trace discord show a sudden birth of quantum correlations. Recall that the initial two qubit state is separable and
the dipole-dipole interaction between the two atoms leads to the generation of non classical correlations. The generated quantum correlations survive over a certain interval of time. After, the amount of quantum correlations
decreases showing a degradation of the generated non classical correlations caused by the environmental effects. It must be noticed that the amount of quantum correlations is more important
when the two atoms get close each other. The interaction between the two atoms  enhances the quantum correlations between the components of the system.  Another important aspect reported in Fig. 4 is the
revival of quantum correlations. Indeed, as it can be seen from Fig. 4, local quantum uncertainty and trace discord increase, reach a maximum and decrease to  vanish after finite time interval. This is
followed by a revival of quantum correlations in the system.  This revival can be explained by the
transfer of correlations from the total system including the environment to the two qubit system.  Comparing the results of Fig. 4 and Fig. 5, one notices that the sudden birth of concurrence is delayed in comparison with local quantum uncertainty and trace quantum discord. This
corroborates the fact that the concurrence as quantum quantifier can not capture the total amount
of quantum correlations existing in a mixed two-qubit system. It is also important to stress that the concurrence does not exhibit the revival phenomena and can not capture the quantum correlation revival.

We also considered the comparison between local quantum uncertainty, trace discord and concurrence for different values of $\gamma$ related to the distance separating the two atoms. The results reported in the figures
6, 7, 8, 9, 10 and 11 show that local quantum uncertainty and trace discord behave almost identically . These figures show also that the concurrence is initially zero for a certain period of time. This means that the evolved states stay separable in this interval but contain quantum
correlations that are not captured by the concurrence. In particular, we noticed that for $\gamma \to 1$, the concurrence is zero. The system of the two atoms stays separable but contains quantum correlations that are not captured by the concurrence.
\begin{figure}[H]
    \centering
    \begin{minipage}[t]{3in}
        \centering
        \includegraphics[width=3.1in]{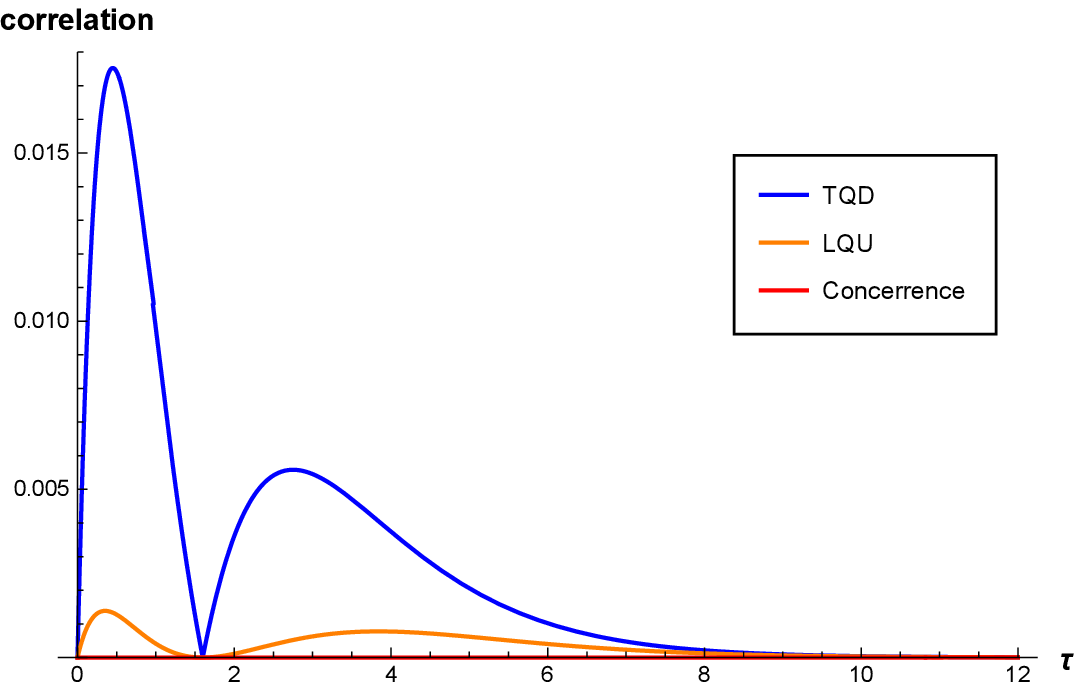}
        {\bf Fig 6:} {\sf Local Quantum Uncertainty, trace Distance Discord and concurrence for $\gamma = 0.1$. }
    \end{minipage}
    \begin{minipage}[t]{3in}
        \centering
        \includegraphics[width=3.1in]{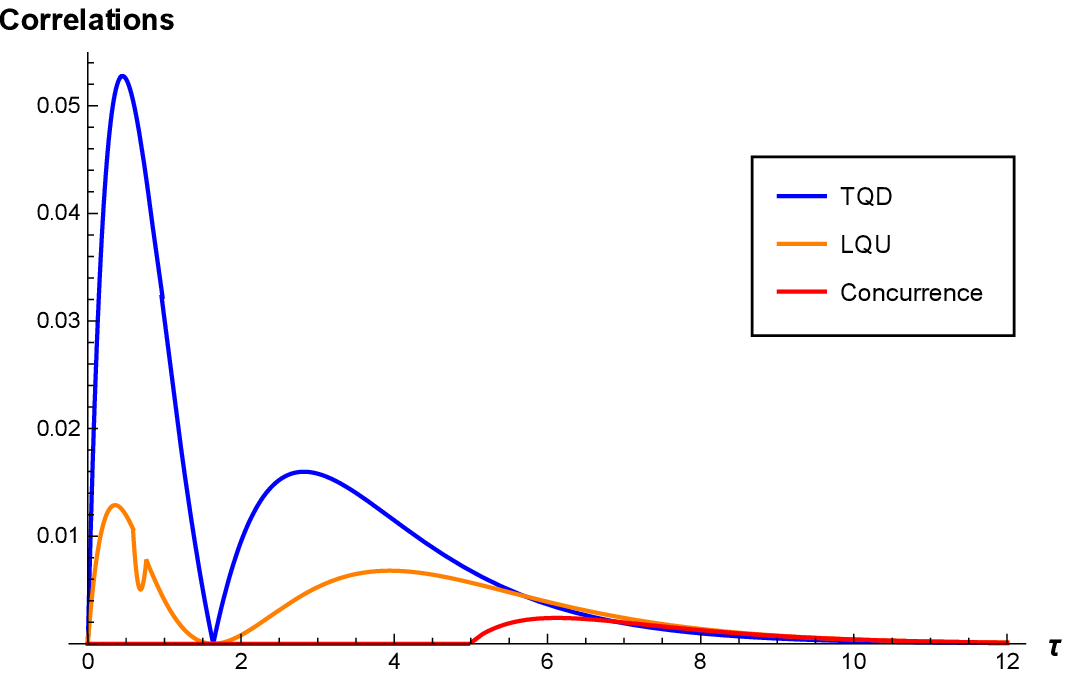}
        {\bf Fig 7:} {\sf Local Quantum Uncertainty, trace Distance Discord and concurrence for $\gamma = 0.3$. }
    \end{minipage}
\end{figure}
\begin{figure}[H]
    \centering
    \begin{minipage}[t]{3in}
        \centering
        \includegraphics[width=3.1in]{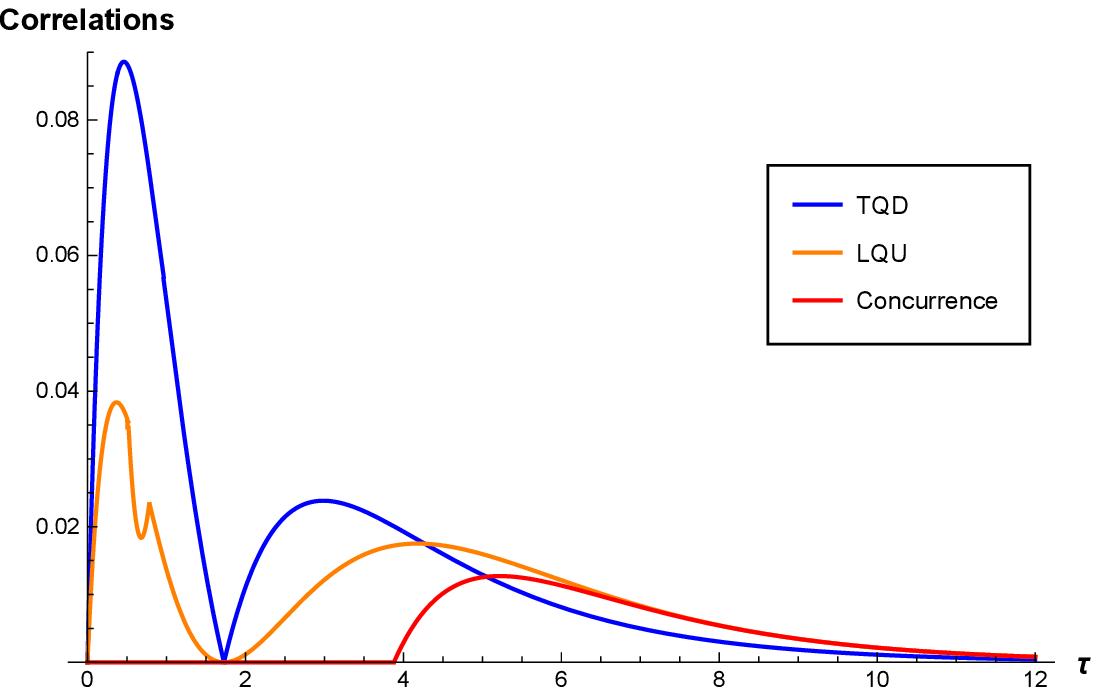}
        {\bf Fig 8:} {\sf Local Quantum Uncertainty, trace Distance Discord and concurrence for $\gamma = 0.5$. }
    \end{minipage}
    \begin{minipage}[t]{3in}
        \centering
        \includegraphics[width=3.1in]{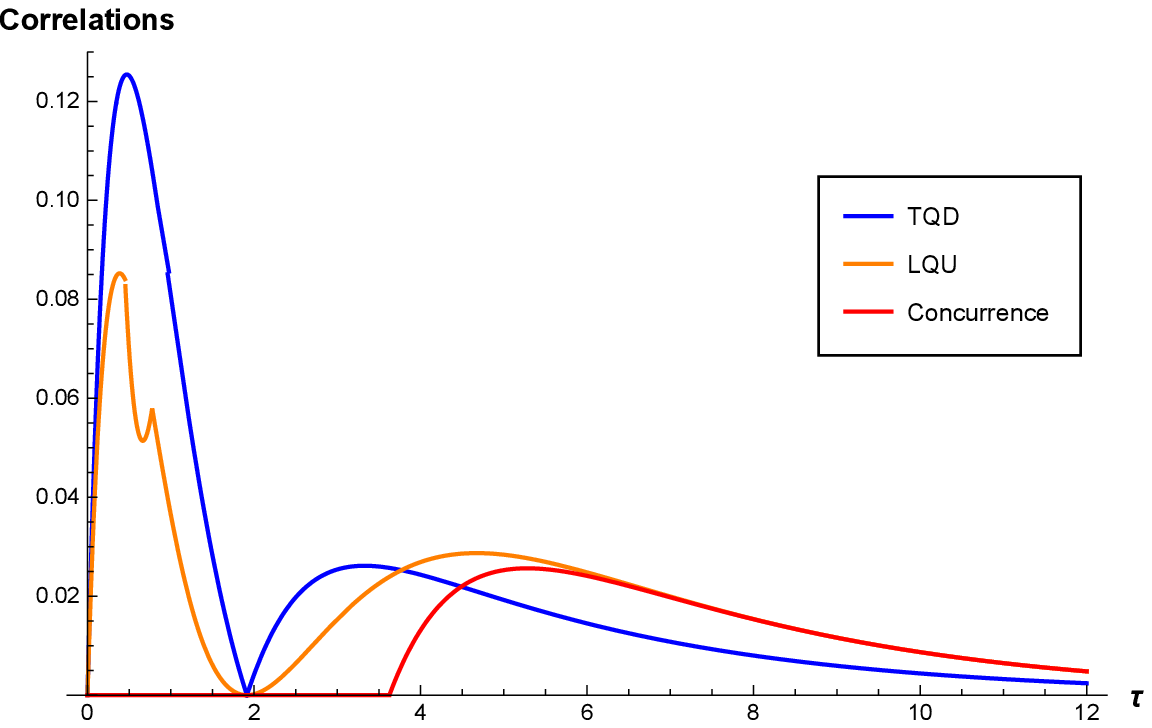}
        {\bf Fig 9:} {\sf Local Quantum Uncertainty, trace Distance Discord and concurrence for $\gamma = 0.7$. }
    \end{minipage}
\end{figure}
\begin{figure}[H]
    \centering
    \begin{minipage}[t]{3in}
        \centering
        \includegraphics[width=3.1in]{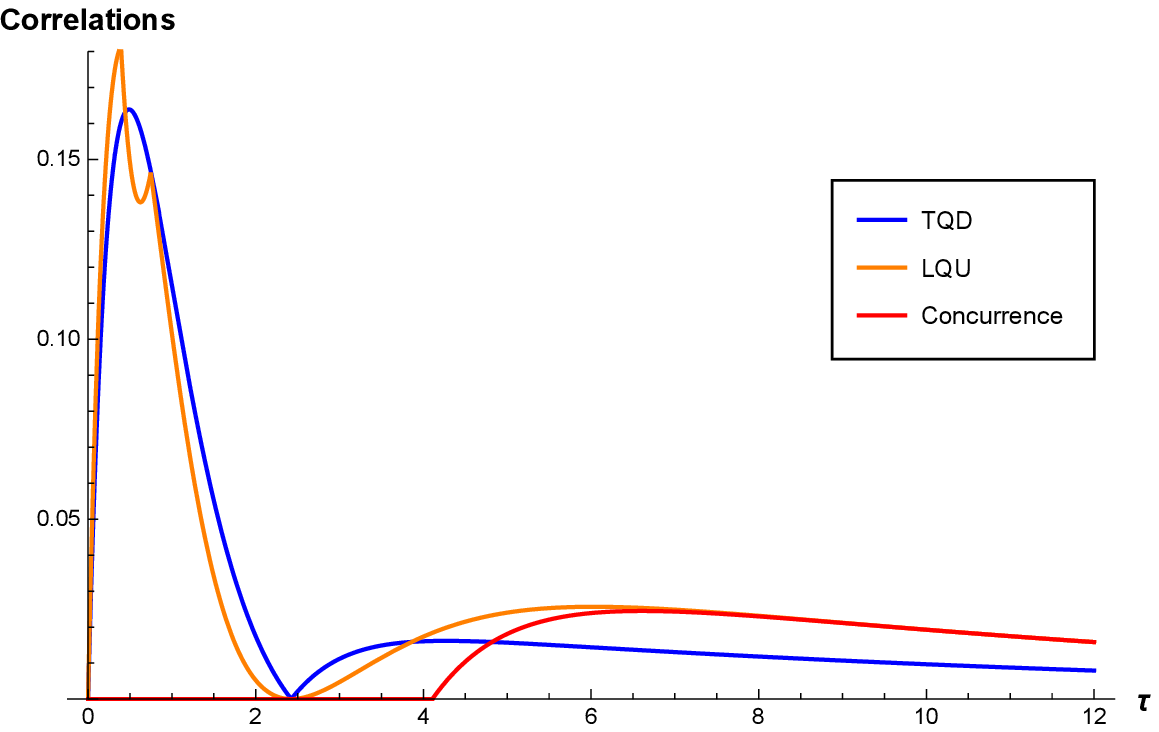}
        {\bf Fig 10:} {\sf Local Quantum Uncertainty, trace Distance Discord and concurrence for $\gamma = 0.9$. }
    \end{minipage}
    \begin{minipage}[t]{3in}
        \centering
        \includegraphics[width=3.1in]{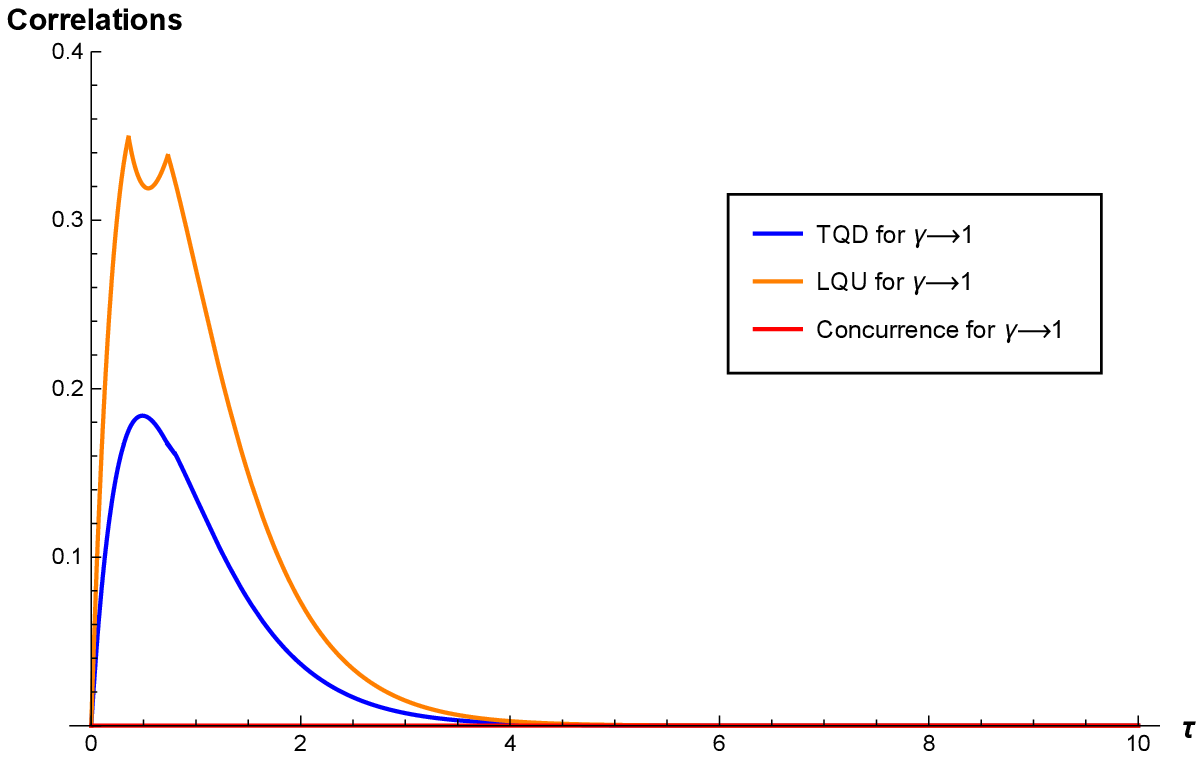}
        {\bf Fig 11:} {\sf Local Quantum Uncertainty, trace Distance Discord and concurrence for $\gamma \longrightarrow 1$. }
    \end{minipage}
\end{figure}

\section{Concluding Remarks}
In this work, we considered the dynamics of quantum correlations in two specific bipartite quantum systems. The first is a two-qubit system coupled to two independent bosonic reservoirs. The
second concerns a  two 2-level atoms interacting with the modes of a quantized radiation field. We used the local quantum uncertainty, trace quantum discord and the concurrence
to investigate the main features of the dynamics of the quantum correlations contained in the system submitted to the environmental effects. In this sense, we have derived the analytical
expression for local quantum uncertainty quantifying quantum correlations in two-qubit $X$ states coupled to two independent bosonic reservoirs. In analyzing
the dynamics of this quantifier, we considered three types of reservoirs :sub-Ohmic, Ohmic, and super-Ohmic. We also compared the local quantum uncertainty and trace quantum discord. In
particular, it has been shown that the local quantum uncertainty does not exhibit the freezing phenomenon which  is observed when one uses the trace norm as quantifier.\\
For a two
2-level atoms interacting
with the bosonic modes  of an electromagnetic field, we analyzed the dynamics of quantum correlations
for different configurations by varying the inter-atomic distance. In this case we have utilized local quantum uncertainty, trace quantum discord and concurrence
to analyze the main characteristics of quantum correlations in this system initially prepared in a separable state. We have noticed that
the evolved system stays for a certain period of time separable (the concurrence is zero) contrarily
to local quantum uncertainty and trace quantum discord which are non vanishing. This corroborates the fact that local quantum uncertainty and trace discord
go beyond the Wootters concurrence and indicates the sudden birth of quantum correlations. This is essentially due to the dipole-dipole interaction
between the two atoms which creates quantum correlations between the components of the system. A second remarkable feature is the
revival phenomenon which reflects the transfer of quantum correlations from the environment to the two qubit system.

\section*{Appendices}
\subsection*{Appendix 1:}
The eigenvalues corresponding to this matrix $\rho$ (\ref{X}) are given by
\begin{equation}
    {{\lambda _1} = \frac{1}{2}{t_1} + \frac{1}{2}\sqrt {{t_1}^2 - 4{d_1}} }, \qquad
    {{\lambda _2} = \frac{1}{2}{t_2} + \frac{1}{2}\sqrt {{t_2}^2 - 4{d_2}} }
\end{equation}
\begin{equation}
    {{\lambda _3} = \frac{1}{2}{t_2} - \frac{1}{2}\sqrt {{t_2}^2 - 4{d_2}} },
\qquad
    {{\lambda _4} = \frac{1}{2}{t_1} - \frac{1}{2}\sqrt {{t_1}^2 - 4{d_1}} }
\end{equation}
where
\begin{equation}
{t_1} = {\rho _{11}} + {\rho _{44}},\quad
        {d_1} = {\rho _{11}}{\rho _{44}} - {\rho _{14}}{\rho _{41}},\quad
        {t_2} = {\rho _{22}} + {\rho _{33}},\quad
        {d_2} = {\rho _{22}}{\rho _{33}} - {\rho _{32}}{\rho _{23}}.
\end{equation}
The square root of the density matrix $\rho $ can be written in terms of the matrix elements $\rho $, in the computational basis, as follows:
\begin{equation}
    \sqrt \rho   = \left( {\begin{array}{*{20}{c}}
            {\frac{{{\rho _{11}} + \sqrt {{d_1}} }}{{\sqrt {{t_1} + 2\sqrt {{d_1}} } }}}&0&0&{\frac{{{\rho _{14}}}}{{\sqrt {{t_1} + 2\sqrt {{d_1}} } }}}\\
            0&{\frac{{{\rho _{22}} + \sqrt {{d_2}} }}{{\sqrt {{t_2} + 2\sqrt {{d_2}} } }}}&{\frac{{{\rho _{23}}}}{{\sqrt {{t_2} + 2\sqrt {{d_2}} } }}}&0\\
            0&{\frac{{{\rho _{32}}}}{{\sqrt {{t_2} + 2\sqrt {{d_2}} } }}}&{\frac{{{\rho _{33}} + \sqrt {{d_2}} }}{{\sqrt {{t_2} + 2\sqrt {{d_2}} } }}}&0\\
            {\frac{{{\rho _{41}}}}{{\sqrt {{t_1} + 2\sqrt {{d_1}} } }}}&0&0&{\frac{{{\rho _{44}} + \sqrt {{d_1}} }}{{\sqrt {{t_1} + 2\sqrt {{d_1}} } }}}
    \end{array}} \right).
\end{equation}
The eigenvalues $\sqrt {{\lambda _1}} $, $\sqrt {{\lambda _2}} $,$\sqrt {{\lambda _3}} $ and $\sqrt {{\lambda _4}} $ of the matrix $\sqrt \rho  $ can be rewritten as:
\begin{equation}
    {\sqrt {{\lambda _1}}  = \frac{1}{2}\sqrt {{t_1} + 2\sqrt {{d_1}} }  + \frac{1}{2}\sqrt {{t_1} - 2\sqrt {{d_1}} } }
\end{equation}
\begin{equation}
    {\sqrt {{\lambda _2}}  = \frac{1}{2}\sqrt {{t_2} + 2\sqrt {{d_2}} }  + \frac{1}{2}\sqrt {{t_2} - 2\sqrt {{d_2}} } }
\end{equation}
\begin{equation}
    {\sqrt {{\lambda _3}}  = \frac{1}{2}\sqrt {{t_2} + 2\sqrt {{d_2}} }  - \frac{1}{2}\sqrt {{t_2} - 2\sqrt {{d_2}} } }
\end{equation}
\begin{equation}
    {\sqrt {{\lambda _4}}  = \frac{1}{2}\sqrt {{t_1} + 2\sqrt {{d_1}} }  - \frac{1}{2}\sqrt {{t_1} - 2\sqrt {{d_1}} } } .
\end{equation}
The matrix $\sqrt \rho  $ is written in the Fano-Bloch
representation as:
\begin{equation}
    \sqrt \rho   = \frac{1}{4}\sum\limits_{\chi ,\delta } {{\mathcal{R}_{\chi \delta }}} {\sigma _\chi } \otimes {\sigma _\delta },
\end{equation}
where $\chi ,\delta  = 0,1,2,3$ and the Fano-Bloch parameters are defined by: ${\mathcal{R}_{\chi \delta }} = {\rm Tr}\left( {\sqrt \rho  {\sigma _\chi } \otimes {\sigma _\delta }} \right)$. The vanishing correlation parameters ${\mathcal{R}_{\chi \delta }}$ are given by:
\begin{equation}
   \mathcal{R}_{00} = \sqrt {{t_1} + 2\sqrt {{d_1}} }  + \sqrt {{t_2} + 2\sqrt {{d_2}} }
\end{equation}
\begin{equation}
    \mathcal{R}_{03} = \frac{1}{2}\frac{{{T_{30}} + {T_{03}}}}{{\sqrt {{t_1} + 2\sqrt {{d_1}} } }} - \frac{1}{2}\frac{{{T_{30}} - {T_{03}}}}{{\sqrt {{t_2} + 2\sqrt {{d_2}} } }}
\end{equation}
\begin{equation}
    \mathcal{R}_{30} = \frac{1}{2}\frac{{{T_{30}} + {T_{03}}}}{{\sqrt {{t_1} + 2\sqrt {{d_1}} } }} + \frac{1}{2}\frac{{{T_{30}} - {T_{03}}}}{{\sqrt {{t_2} + 2\sqrt {{d_2}} } }}
\end{equation}
\begin{equation}
    \mathcal{R}_{11} = \frac{1}{2}\frac{{{T_{11}} + {T_{22}}}}{{\sqrt {{t_2} + 2\sqrt {{d_2}} } }} + \frac{1}{2}\frac{{{T_{11}} - {T_{22}}}}{{\sqrt {{t_1} + 2\sqrt {{d_1}} } }}
\end{equation}
\begin{equation}
    \mathcal{R}_{12} = \frac{1}{2}\frac{{{T_{12}} - {T_{21}}}}{{\sqrt {{t_2} + 2\sqrt {{d_2}} } }} + \frac{1}{2}\frac{{{T_{12}} + {T_{21}}}}{{\sqrt {{t_1} + 2\sqrt {{d_1}} } }}
\end{equation}
\begin{equation}
    \mathcal{R}_{12} = \frac{1}{2}\frac{{{T_{12}} - {T_{21}}}}{{\sqrt {{t_2} + 2\sqrt {{d_2}} } }} + \frac{1}{2}\frac{{{T_{12}} + {T_{21}}}}{{\sqrt {{t_1} + 2\sqrt {{d_1}} } }}
\end{equation}
\begin{equation}
    \mathcal{R}_{21} = \frac{1}{2}\frac{{{T_{12}} + {T_{21}}}}{{\sqrt {{t_1} + 2\sqrt {{d_1}} } }} - \frac{1}{2}\frac{{{T_{12}} - {T_{21}}}}{{\sqrt {{t_2} + 2\sqrt {{d_2}} } }}
\end{equation}
\begin{equation}
    \mathcal{R}_{22} = \frac{1}{2}\frac{{{T_{11}} + {T_{22}}}}{{\sqrt {{t_2} + 2\sqrt {{d_2}} } }} - \frac{1}{2}\frac{{{T_{11}} - {T_{22}}}}{{\sqrt {{t_1} + 2\sqrt {{d_1}} } }}
\end{equation}
\begin{equation}
    \mathcal{R}_{33} = \sqrt {{t_1} + 2\sqrt {{d_1}} }  - \sqrt {{t_2} + 2\sqrt {{d_2}} }  .
\end{equation}
\subsection*{Appendix 2:}
In this appendix, we give the expressions of the $\gamma \left( t \right)$ function for sub-ohmic, ohmic and super-ohmic reservoirs \cite{liu}.
\subsubsection*{(\emph{i}) Sub-Ohmic reservoirs:}
The sub-Ohmic regime corresponds to the situation where  $0 < s < 1$. In this case, the function $\gamma \left( t \right)$ reads as \cite{liu}
{\footnotesize{\begin{align*}
            \gamma \left( t \right) &= \frac{{2\lambda \Gamma \left( s \right)}}{{s - 1}}\left\{ {1 - {{\left( {1 - {\Omega ^2}{t^2}} \right)}^{{\raise0.7ex\hbox{${\left( {1 - s} \right)}$} \!\mathord{\left/
                                {\vphantom {{\left( {1 - s} \right)} 2}}\right.\kern-\nulldelimiterspace}
                            \!\lower0.7ex\hbox{$2$}}}}\cos \left[ {\left( {s - 1} \right)\arctan \left( {\Omega t} \right)} \right]} \right\}\\
            &+ \frac{{4\lambda \Gamma \left( s \right)}}{{s - 1}}{\sum\limits_{m = 1}^\infty  {\left( {1 + m\Omega \beta } \right)} ^{1 - s}} \times \left\{ {1 - {{\left[ {1 + {{\left( {\frac{{\Omega t}}{{1 + m\Omega \beta }}} \right)}^2}} \right]}^{{\raise0.7ex\hbox{${\left( {1 - s} \right)}$} \!\mathord{\left/
                                {\vphantom {{\left( {1 - s} \right)} 2}}\right.\kern-\nulldelimiterspace}
                            \!\lower0.7ex\hbox{$2$}}}}\cos \left[ {\left( {s - 1} \right)\arctan \left( {\frac{{\Omega t}}{{1 + m\Omega \beta }}} \right)} \right]} \right\},
\end{align*}}}
where ${\Gamma \left( s \right)}$ denotes the Gamma function. To study the evolution of the local quantum uncertainty, one should  calculate the time derivative of the function $\gamma \left( t \right)$ . It is given by
\begin{equation}
    \scriptsize{\frac{{d\gamma \left( t \right)}}{{dt}} = 2\lambda \Gamma \left( s \right)\Omega \left\{ {{{\left( {1 + {\Omega ^2}{t^2}} \right)}^{{{ - s} \mathord{\left/
                            {\vphantom {{ - s} 2}} \right.
                            \kern-\nulldelimiterspace} 2}}}\sin \left[ {s\arctan \left( {\Omega t} \right)} \right] + 2{{\sum\limits_{m = 1}^\infty  {\left[ {{{\left( {1 + m\Omega \beta } \right)}^2} + {\Omega ^2}{t^2}} \right]} }^{{{ - s} \mathord{\left/
                            {\vphantom {{ - s} 2}} \right.
                            \kern-\nulldelimiterspace} 2}}}\sin \left[ {s\arctan \left( {\frac{{\Omega t}}{{1 + m\Omega \beta }}} \right)} \right]} \right\}}
\end{equation}
\subsubsection*{(\emph{ii}) Ohmic reservoirs:}
For reservoirs of  Ohmic type $(s = 1)$, the low-frequency behavior of the function $J\left( \omega  \right)$ is linear in term of $\omega $. The equation (\ref{gama}) becomes \cite{liu}
\begin{equation}
    \gamma \left( t \right) = \lambda \left[ {\ln \left( {1 + {\Omega ^2}{t^2}} \right) + 4\ln \Gamma \left( {1 + \frac{1}{{\Omega \beta }}} \right) - 2\ln {{\left| {\Gamma \left( {1 + \frac{1}{{\Omega \beta }} + i\frac{t}{\beta }} \right)} \right|}^2}} \right]
\end{equation}
The time derivative of the function $\gamma \left( t \right)$  is given by
\begin{equation}
    \frac{{d\gamma \left( t \right)}}{{dt}} = 2\lambda {\Omega ^2}t\left[ {\frac{1}{{1 + {\Omega ^2}{t^2}}} + \sum\limits_{m = 1}^\infty  {\frac{2}{{{{\left( {1 + m\Omega \beta } \right)}^2} + {\Omega ^2}{t^2}}}} } \right].
\end{equation}
\subsubsection*{(\emph{iii}) Super-Ohmic reservoirs:}
For super-Ohmic reservoirs $(s > 1)$, the function  $\gamma \left( t \right)$ reads as \cite{liu}
\begin{align*}
    \gamma \left( t \right) &= 2\lambda \Gamma \left( {s - 1} \right)\left\{ {1 - {{\left( {1 + {\Omega ^2}{t^2}} \right)}^{{{1 - s} \mathord{\left/
                        {\vphantom {{1 - s} 2}} \right.
                        \kern-\nulldelimiterspace} 2}}}\cos \left[ {\left( {s - 1} \right)\arctan \left( {\Omega t} \right)} \right]} \right\}\\ &+ 4\lambda \Gamma \left( {s - 1} \right)\sum\limits_{m = 1}^\infty  {{{\left( {1 + m\Omega \beta } \right)}^{1 - s}}\left\{ {1 - {{\left[ {1 + {{\left( {\frac{{\Omega t}}{{1 + m\Omega \beta }}} \right)}^2}} \right]}^{{{\left( {1 - s} \right)} \mathord{\left/
                            {\vphantom {{\left( {1 - s} \right)} 2}} \right.
                            \kern-\nulldelimiterspace} 2}}}\cos \left[ {\left( {s - 1} \right)\arctan \left( {\frac{{\Omega t}}{{1 + m\Omega \beta }}} \right)} \right]} \right\}},
\end{align*}
and the derivative of $\gamma \left( t \right)$ with respect to time $t$ is given by
\begin{align*}
    \frac{{d\gamma \left( t \right)}}{{dt}}&= 2\lambda \Omega \Gamma \left( s \right){\left( {1 + {\Omega ^2}{t^2}} \right)^{{{ - s} \mathord{\left/
                    {\vphantom {{ - s} 2}} \right.
                    \kern-\nulldelimiterspace} 2}}}\sin \left[ {s\arctan \left( {\Omega t} \right)} \right]\\
    &+ 4\lambda \Omega \Gamma \left( s \right){\sum\limits_{m = 1}^\infty  {\left[ {{{\left( {1 + m\Omega \beta } \right)}^2} + {\Omega ^2}{t^2}} \right]} ^{ - {s \mathord{\left/
                    {\vphantom {s 2}} \right.
                    \kern-\nulldelimiterspace} 2}}}\sin \left[ {s\arctan \left( {\frac{{\Omega t}}{{1 + m\Omega \beta }}} \right)} \right].
\end{align*}
Due to the complicated expression of $\gamma \left( t \right)$ and to simplify our numerical calculations, we have chosen the situation where $1 < s \le 2$ and the temperature equals zero. In  this case case, the function $\gamma \left( t \right)$ becomes
\begin{equation}
    \gamma \left( t \right) = 2\lambda \Gamma \left( {s - 1} \right)\left\{ {1 - {{\left( {1 + {\Omega ^2}{t^2}} \right)}^{{{\left( {1 - s} \right)} \mathord{\left/
                        {\vphantom {{\left( {1 - s} \right)} 2}} \right.
                        \kern-\nulldelimiterspace} 2}}}\cos \left[ {\left( {s - 1} \right)\arctan \left( {\Omega t} \right)} \right]} \right\}.
\end{equation}

\end{document}